\documentclass[traditabstract]{aa} 
\usepackage{graphicx}
\usepackage[varg]{txfonts}  
\usepackage{natbib}
\usepackage{tikz}
\graphicspath{{figures/}}
\pdfpageattr{/Group <</S /Transparency /I true /CS /DeviceRGB>>}
\newcommand{\e}{{\rm e}}

\newcommand{\Spla}{{{\cal S}_p}}

\newcommand{\Fstar}{{{\cal F}_{\rm star}}}

\newcommand{\Fpla}{{{\cal F}_{\rm pla}}}
\newcommand{\Ftrans}{{{\cal F}_{\rm transit}}}
\newcommand{\Gfit}{{{\cal G}_{\sigma}}}
\newcommand{\Gfitb}{{{\cal G}_{\sigma_0}}}
\newcommand{\va}{{v_{\rm mean}}}
\newcommand{\vb}{{v_{\rm H10}}}
\newcommand{\vc}{{v_{\rm CCF}}}
\newcommand{\vd}{{v_{\rm iodine}}}
\newcommand{\vx}{{\bar v}}

\newcommand{\moy}[2]{\left\langle{#2}\right\rangle_{#1}}
\def\crm{\cr\noalign{\medskip}}

\def\m@th{\mathsurround=0pt}
\def\EQM#1{\vcenter{\normalbaselines\m@th
    \ialign{${\displaystyle ##}$\hfil&&\ ${\displaystyle ##}$\hfil\crcr
    \mathstrut\crcr\noalign{\kern-\baselineskip}
    \noalign{\smallskip}
    #1\crcr\mathstrut\crcr\noalign{\kern-\baselineskip}}}}

\newcommand{\be}{\begin{equation}}
\newcommand{\ee}{\end{equation}}

\def\Dron#1#2{\frac{\partial#1}{\partial#2}}

\newcommand{\bpm}{\left(\begin{array}{c}}
\newcommand{\epm}{\end{array}\right)}
\newcommand{\figexpl}{
\begin{figure*}
\begin{center}
\includegraphics[width=\linewidth]{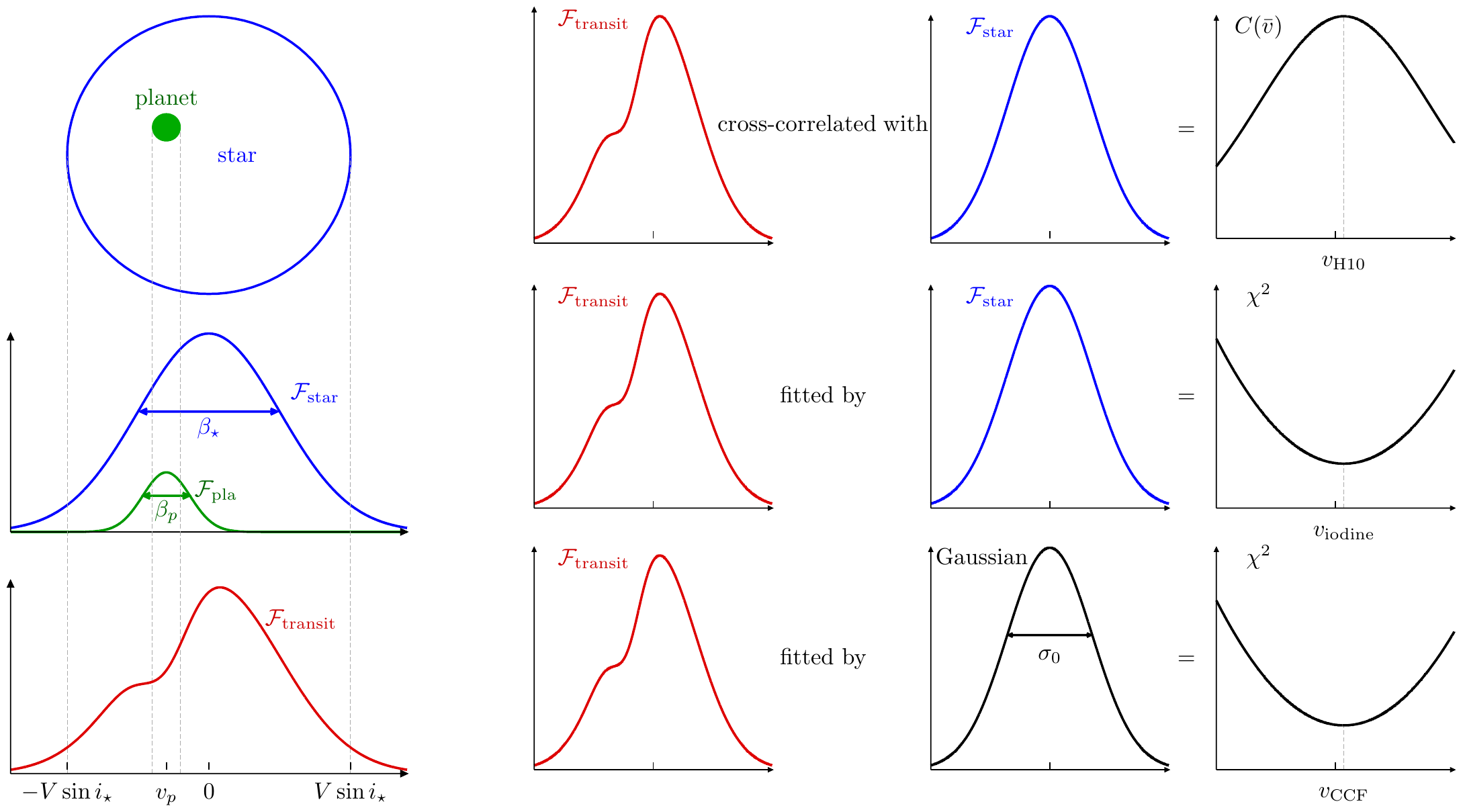}
\caption{\label{fig.explanation} Simplified illustration of different methods to
compute the Rossiter-McLaughlin effect. $v_{\rm H10}$, $v_{\rm iodine}$,
and $v_{\rm CCF}$ represent the result of the hypothesis made in
\citet{Hirano_etal_ApJ_2010}, the result of the data reduction done with
the iodine cell technique, and the result of the data reduction done
with the CCF technique, respectively.}
\end{center}
\end{figure*}
}
\newcommand{\figCoord}{
\begin{figure}
\begin{center}
\includegraphics[width=\linewidth]{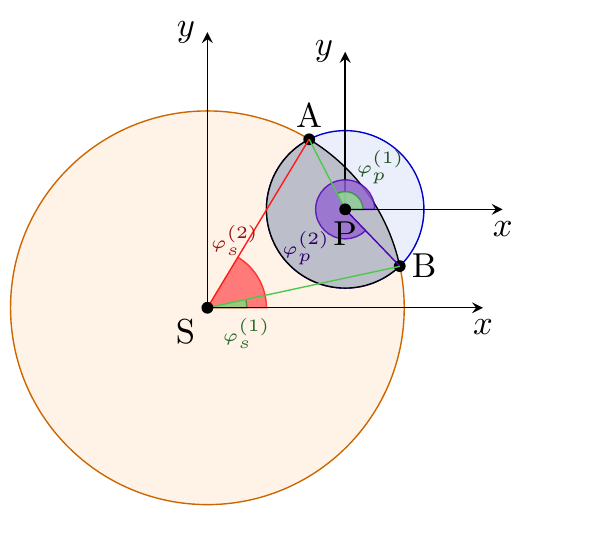}
\caption{\label{fig.coord} Definition of the angles $\varphi_p^{(1)}$, 
$\varphi_p^{(2)}$, $\varphi_s^{(1)}$, and $\varphi_s^{(2)}$ during partial 
transit. The large circle centered on S represents the star, and the
smaller one, centered on P, is the planet.}
\end{center}
\end{figure}
}
\newcommand{\figflux}{
\begin{figure}
\begin{center}
\includegraphics[width=\linewidth]{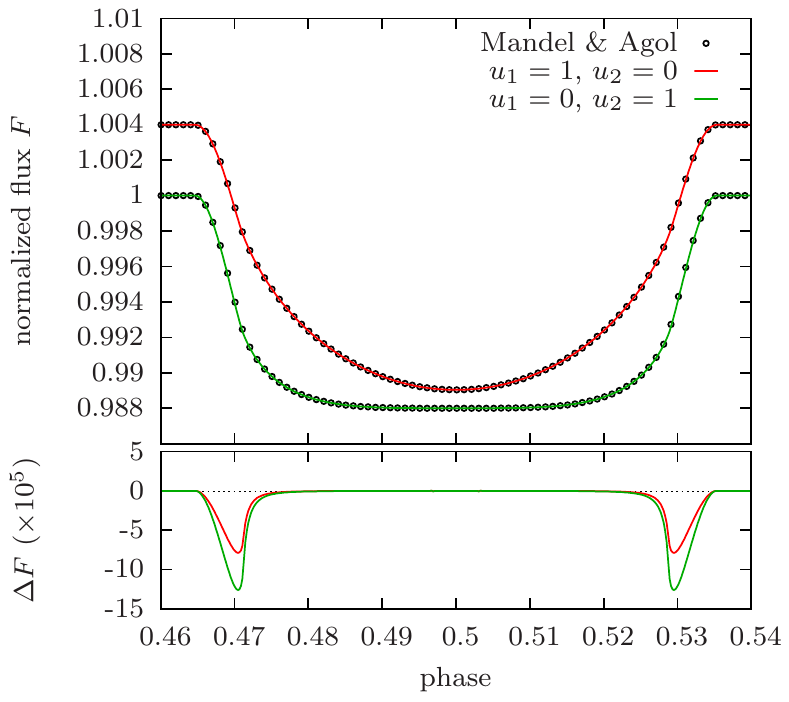}
\caption{\label{fig.flux} Transit light curves for $r=0.1$ and quadratic
limb-darkening with two sets of coefficients $u_1$, $u_2$. The solid
curves in red and in green are obtained from the approximation
(\ref{eq.fsimp}), while the black open circles are computed using the
routine of \citet{Mandel_Agol_ApJ_2002}.}
\end{center}
\end{figure}
}
\newcommand{\figplanetprofile}{
\begin{figure}
\begin{center}
\includegraphics[width=\linewidth]{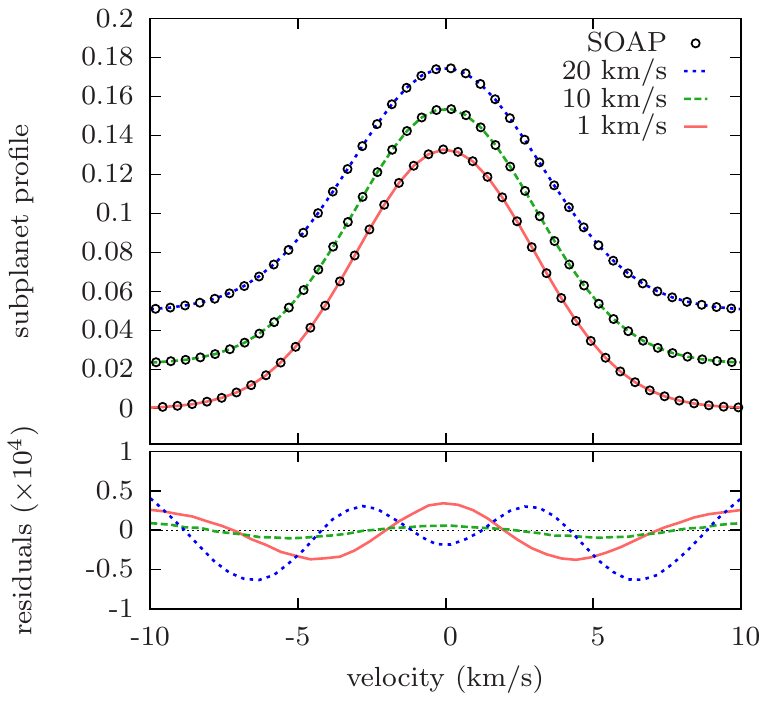}
\caption{\label{fig.planetprofile} Example of subplanet line profiles
obtained with SOAP-T (circles), compared with Gaussian profiles (curves) for
different stellar $V\sin i_\star$.}
\end{center}
\end{figure}
}
\newcommand{\figvp}{
\begin{figure}
\begin{center}
\includegraphics[width=\linewidth]{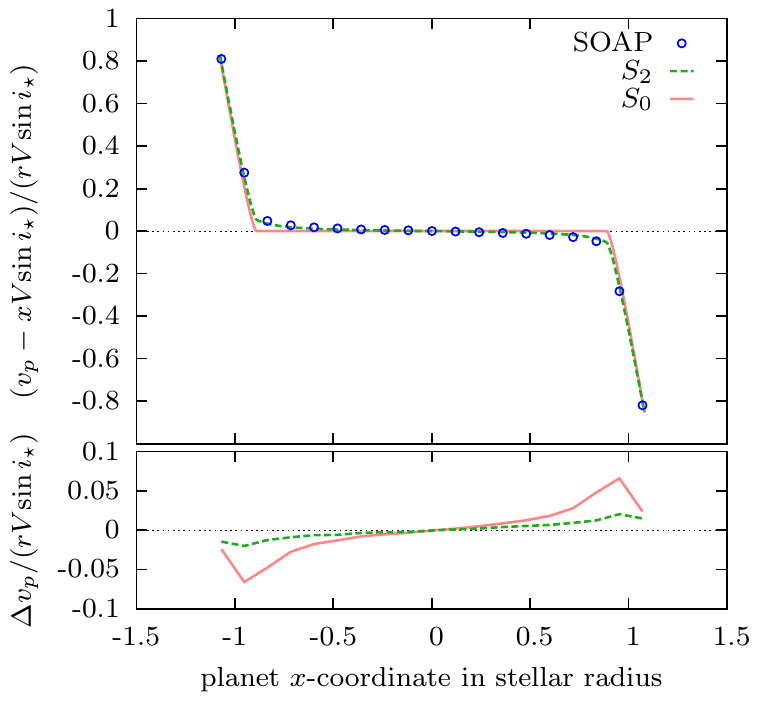}
\caption{\label{fig.vp} Subplanet velocity $v_p$ produced with SOAP-T
(blue points), approximation $S_0$ assuming uniform intensity below the
disk of the planet (red curve), and approximation $S_2$ taking the
second derivatives of the stellar surface brightness into account,
Eq.~(\ref{eq.vp}) (green curve).}
\end{center}
\end{figure}
}
\newcommand{\figbetap}{
\begin{figure}
\begin{center}
\includegraphics[width=\linewidth]{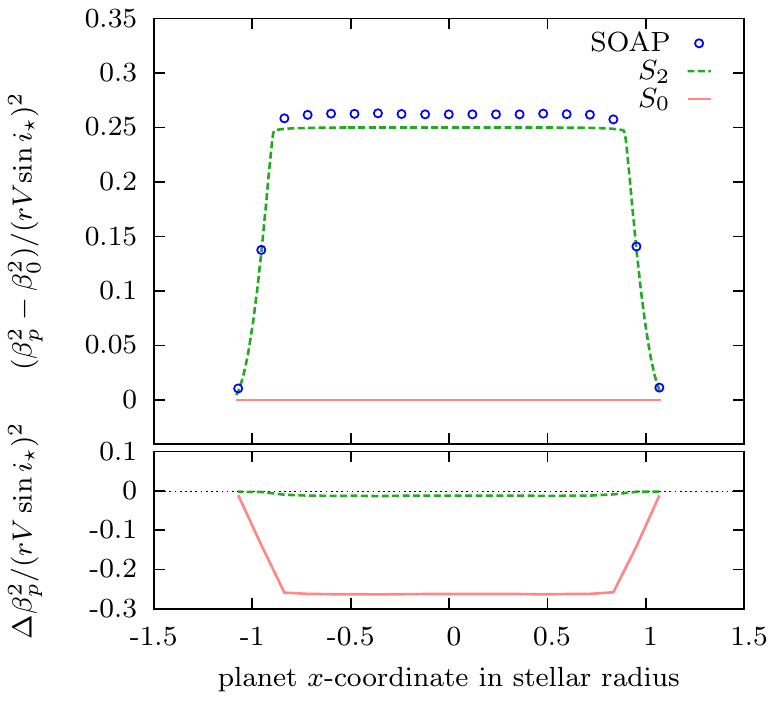}
\caption{\label{fig.betap} Subplanet dispersion $\beta_p$ produced with
SOAP-T (blue points), approximation $S_0$ (red curve), and approximation
$S_2$, Eq.~(\ref{eq.betap}), (green curve).}
\end{center}
\end{figure}
}
\newcommand{\figRM}{
\begin{figure}
\begin{center}
\includegraphics[width=\linewidth]{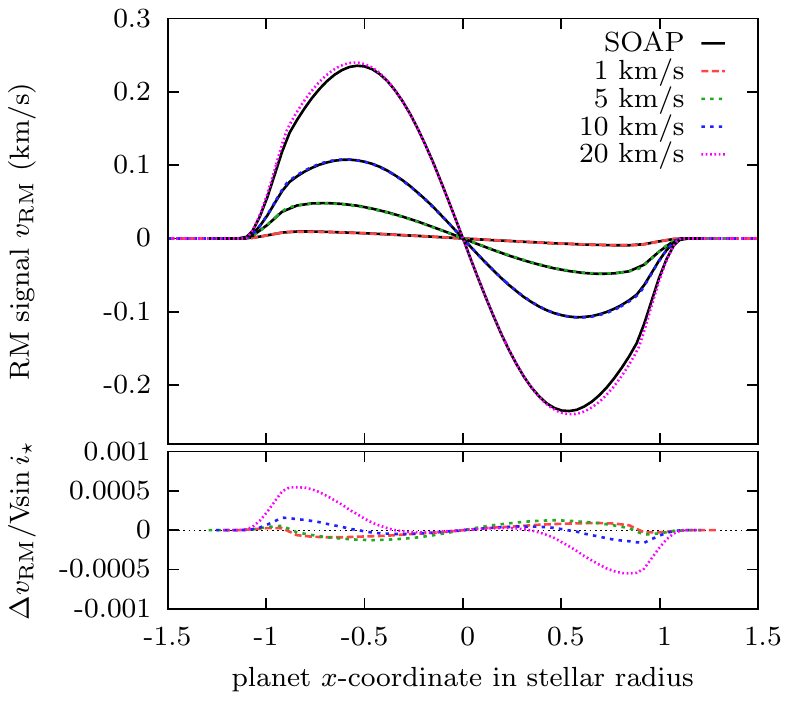}
\caption{\label{fig.RM} RM signals produced with SOAP-T (solid black curves) for
different $V\sin i_\star$, and results of $\vc$ (\ref{eq.solexact})
(different dashed curves).}
\end{center}
\end{figure}
}
\newcommand{\figLines}{
\begin{figure*}
\begin{center}
\includegraphics[width=0.8\linewidth]{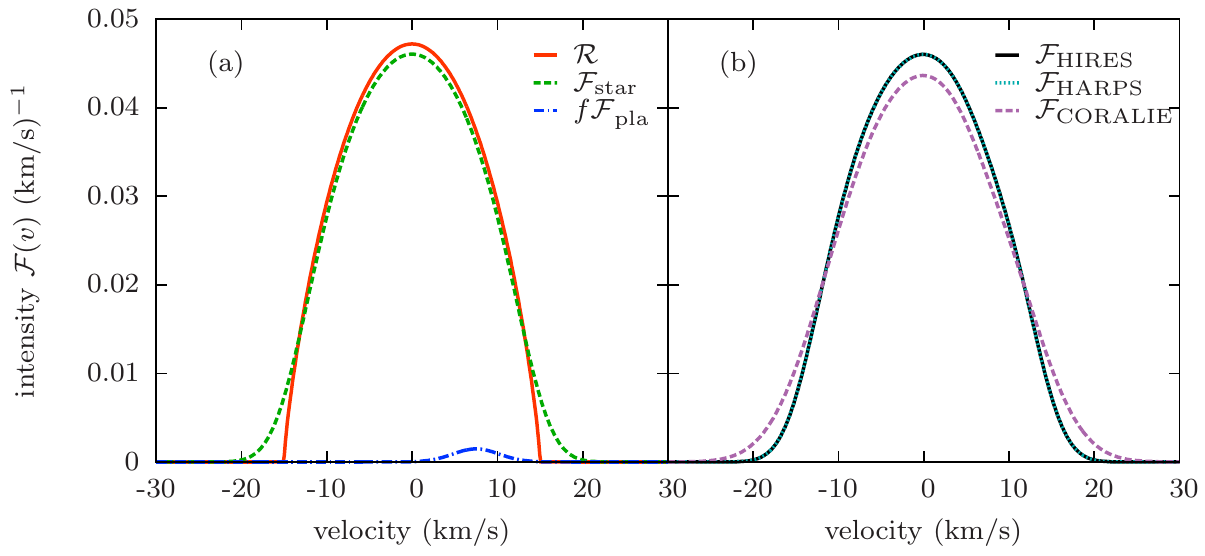}
\caption{\label{fig.linetest} Simple models of line profiles. (a)
Rotation kernel ${\cal R}(v)$ with $V\sin i_\star=15$ km.s$^{-1}$ in
solid red, stellar line profile assuming $\beta_0=2.6$ km.s$^{-1}$ in
dashed green, subplanet line profile $\Fpla(v)$ with $\beta_p=2.71$
km.s$^{-1}$ in dash-dotted blue. (b) $\Ftrans=\Fstar-f\Fpla$ modeling an
average line profile observed with HIRES in solid black, and a CCF
observed with HARPS in dotted cyan. The same with $\beta_0=4.5$
km.s$^{-1}$ and $\beta_p=4.56$ km.s$^{-1}$ represents a CCF observed by
CORALIE, in dashed violet.}
\end{center}
\end{figure*}
}
\newcommand{\figRMcomp}{
\begin{figure*}
\begin{center}
\includegraphics[width=0.8\linewidth]{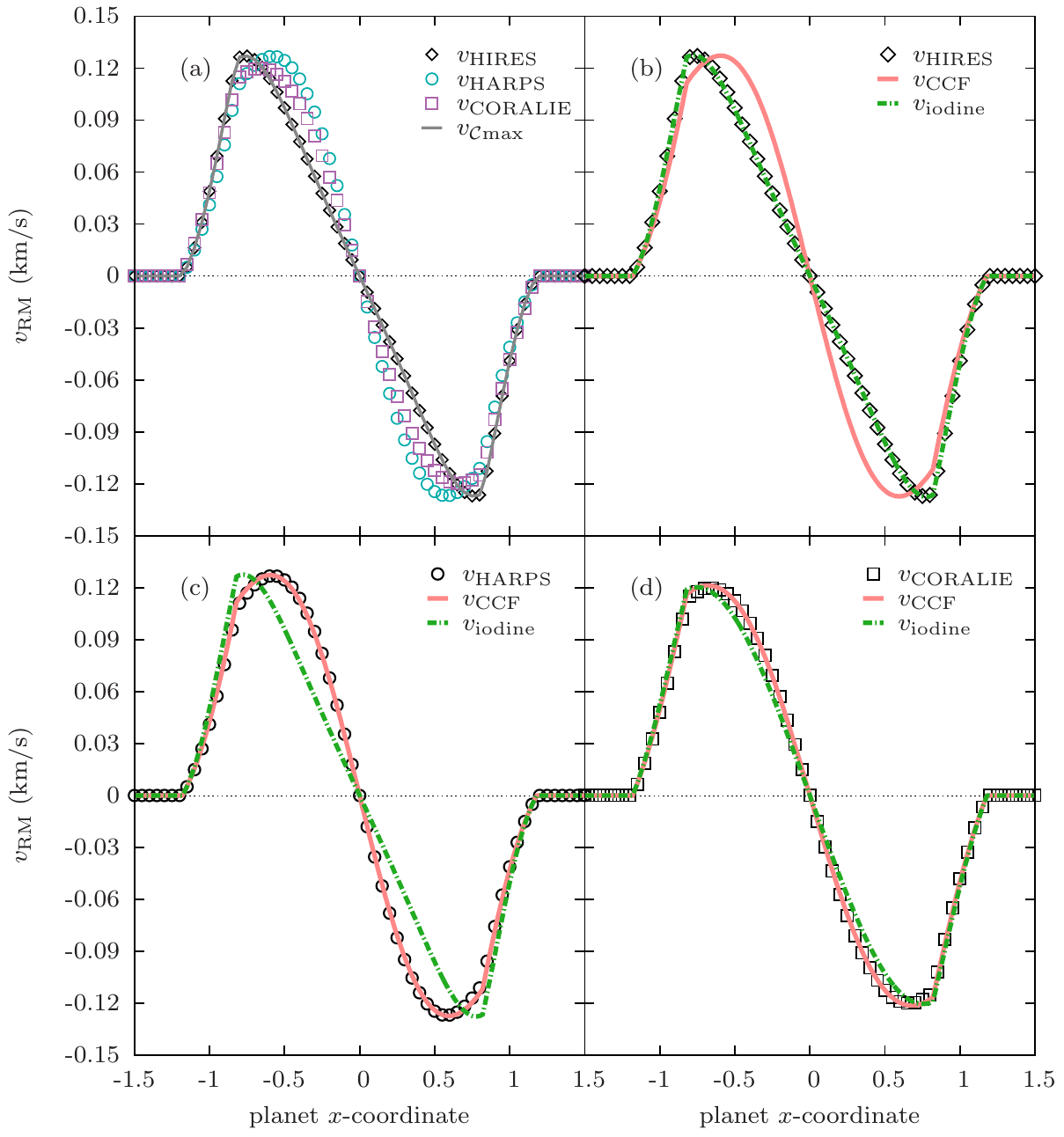}
\caption{\label{fig.RMcomp} Comparison of a simulated RM signal observed
by different techniques and/or instruments. We used the line profiles of 
Fig.~\ref{fig.linetest}b. The open diamonds, circles and squares represent 
the RM signal obtained numerically with the iodine cell technique on the
HIRES line profile, and with the Gaussian fit to the HARPS, or CORALIE,
CCFs, respectively. In (a) The gray curve corresponds to the numerical
maximization of the cross-correlation $C(\vx)$ of the HIRES profiles
inside and outside transits. In (b), (c), and (d), the numerical RM
signal computed on the HIRES, HARPS, and CORALIE line profiles,
respectively, are compared with the analytical formulas $\vc$
(\ref{eq.solexact}) in solid red and $\vd$ (\ref{eq.vdconv}) in
dash-dotted green.}
\end{center}
\end{figure*}
}
\newcommand{\figBiasa}{
\begin{figure}
\begin{center}
\includegraphics[width=\linewidth]{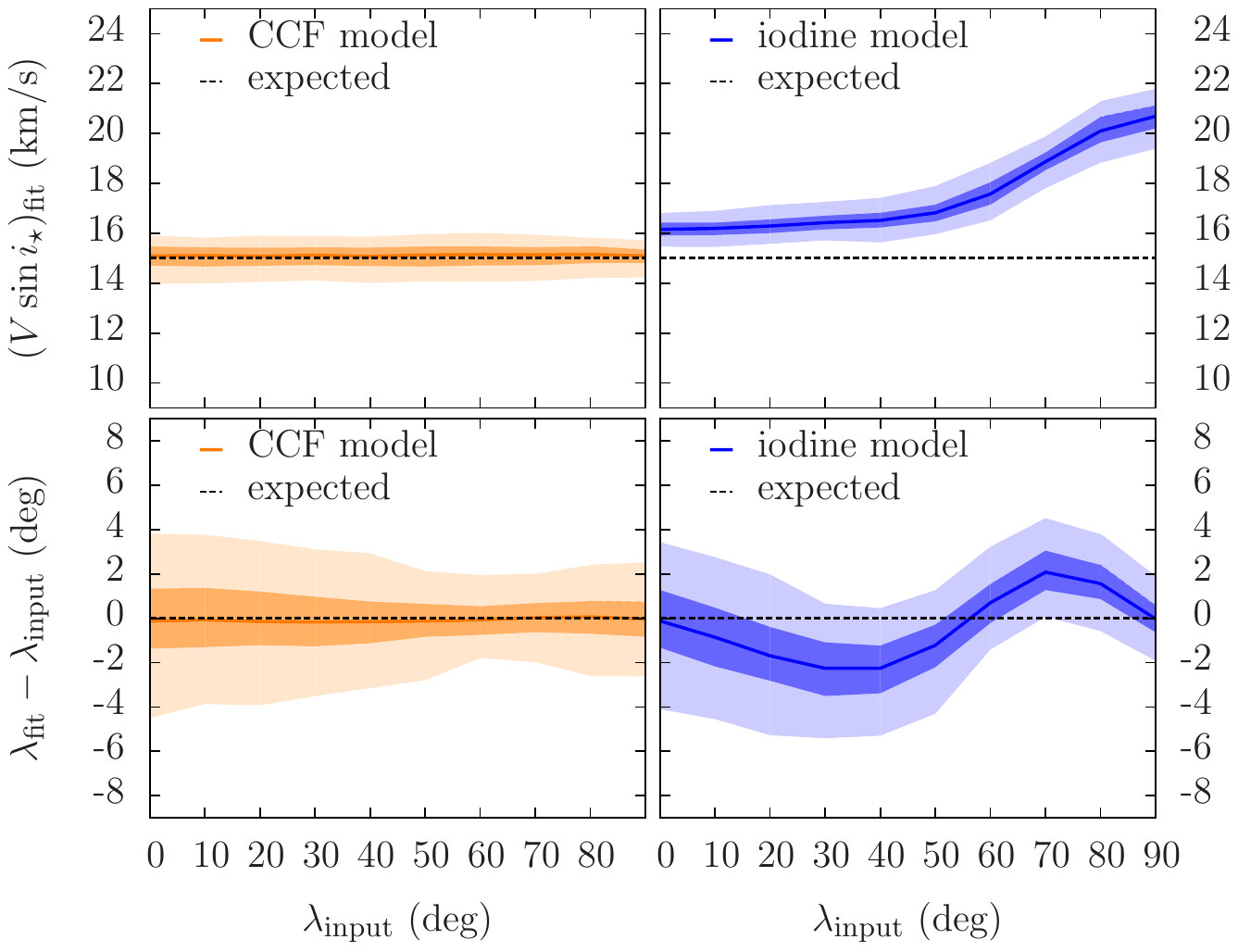}
\caption{\label{fig.bias_CCF} Results of the fits with two different models of
mock data generated by the CCF technique formulas. In each panel, the area
with the lightest color represents the two-sigma limit, the darkest color
is the one-sigma threshold, and the curve is the best value.}
\end{center}
\end{figure}
}
\newcommand{\figBiasb}{
\begin{figure}
\begin{center}
\includegraphics[width=\linewidth]{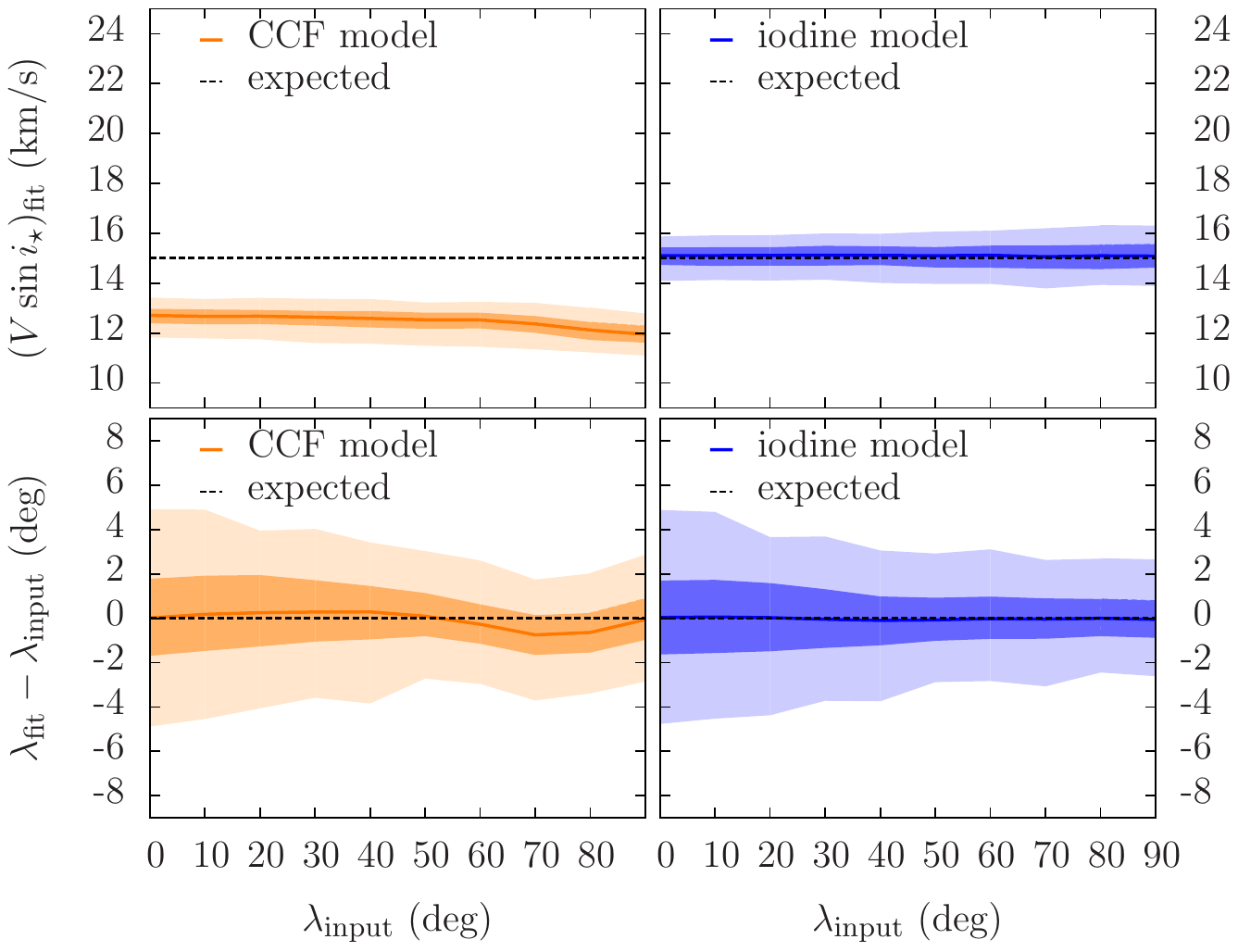}
\caption{\label{fig.bias_iodine} Same as Fig.~\ref{fig.bias_CCF} but for
data simulated with the iodine cell technique formulas.}
\end{center}
\end{figure}
}
\newcommand{\tabHess}{
\begin{table}
\caption{\label{tab.Hessian} Second derivatives of $x^nI_\alpha(x,y)$, Eq.~(\ref{eq.Ialpha}).}
\begin{center}
\renewcommand{\arraystretch}{1.0}
\begin{tabular}{l} \hline\hline
\\
$\displaystyle
\EQM{
H_{xx}^{(n)} =& n(n-1)x^{n-2}I_\alpha(x,y) 
              -n(n+1)\,\alpha x^n I_{\alpha-2}(x,y)
\crm &
              +\alpha(\alpha-2)x^{n+2} I_{\alpha-4}(x,y)
}
$ \\ \\
$\displaystyle
H_{yy}^{(n)} = - n\alpha x^n I_{\alpha-2}(x,y)
               +\alpha(\alpha-2)x^n y^2 I_{\alpha-4}(x,y)
$ \\ \\
$\displaystyle
H_{xy}^{(n)} = - n\alpha x^{n-1} y I_{\alpha-2}(x,y)
               +\alpha(\alpha-2)x^{n+1} y^2 I_{\alpha-4}(x,y)
$ \\ \\
\hline
\end{tabular}
\end{center}
\end{table}
}
\newcommand{\tabintegFij}{
\begin{table*}
\caption{\label{tab.integFij} Integrals $F_{ij}$ (\ref{eq.Fij}) of the
field vectors ${\vec f}_{ij}(x,y)$ along a circular arc centered on $(x_0,y_0)$ with radius $r$.}
\begin{center}
\renewcommand{\arraystretch}{1.0}
\begin{tabular}{l} \hline\hline
\\
$\displaystyle
F_{0,0}(\varphi) = \frac{1}{2}r(r\varphi+x_0\sin\varphi-y_0\cos\varphi)
$ \\ \\
$\displaystyle
F_{1,0}(\varphi) =-x_0y_0r\cos\varphi 
         -\frac{1}{2}y_0 r^2 \cos^2\varphi
         +\frac{1}{4}x_0 r^2 (2\varphi-\sin2\varphi)
         +\frac{1}{12}r^3(3\sin\varphi-\sin3\varphi)
$\\ \\
$\displaystyle
F_{0,1}(\varphi) = x_0y_0r\sin\varphi
         +\frac{1}{2}x_0 r^2 \sin^2\varphi
         +\frac{1}{4}y_0 r^2 (2\varphi+\sin2\varphi)
         -\frac{1}{12}r^3(3\cos\varphi+\cos3\varphi)
$\\ \\
$\displaystyle
F_{2,0}(\varphi) = -x_0^2y_0r\cos\varphi
          -x_0y_0r^2\cos^2\varphi
          +\frac{1}{4}x_0^2r^2(2\varphi-\sin2\varphi)
          -\frac{1}{12}y_0r^3(3\cos\varphi+\cos3\varphi)
          +\frac{1}{6}x_0r^3(3\sin\varphi-\sin3\varphi)
          +\frac{1}{32}r^4(4\varphi-\sin4\varphi)
$\\ \\
$\displaystyle
F_{0,2}(\varphi) =  x_0y_0^2r\sin\varphi
          +x_0y_0r^2\sin^2\varphi
          +\frac{1}{4}y_0^2r^2(2\varphi+\sin2\varphi)
          +\frac{1}{12}x_0r^3(3\sin\varphi-\sin3\varphi)
          -\frac{1}{6}y_0r^3(3\cos\varphi+\cos3\varphi)
          +\frac{1}{32}r^4(4\varphi-\sin4\varphi)
$\\ \\
$\displaystyle
F_{1,1}(\varphi) =  \frac{1}{4}x_0y_0r(2r\varphi+x_0\sin\varphi-y_0\cos\varphi)
          +\frac{1}{8}x_0^2r^2\sin^2\varphi
          -\frac{1}{8}(y_0^2+r^2)r^2\cos^2\varphi
          +\frac{1}{48}y_0r^3(15\sin\varphi-\sin3\varphi)
          -\frac{1}{48}x_0r^3(15\cos\varphi+\cos3\varphi)
$\\ \\
\hline
\end{tabular}
\end{center}
\end{table*}
}
\begin{document}
   \title{New analytical expressions of the Rossiter-McLaughlin effect
adapted to different observation techniques}
   \author{Gwena\"el Bou\'e\inst{1,2,3}
           \and
           Marco Montalto\inst{1}
           \and
           Isabelle Boisse\inst{1}
           \and
           Mahmoudreza Oshagh\inst{1,4}
           \and
           Nuno C. Santos\inst{1,4}
          }
   \institute{Centro de Astrof\'isica, Universidade do Porto, Rua das
              Estrelas, 4150-762 Porto, Portugal
         \and
             Astronomie et Syst\`emes Dynamiques, IMCCE-CNRS UMR8028,
             Observatoire de Paris, UPMC, 77 Av. Denfert-Rochereau, 
             75014 Paris, France
         \and
             Department of Astronomy and Astrophysics, University of
             Chicago, 5640 South Ellis Avenue, Chicago, IL 60637, USA\\
             \email{boue@oddjob.uchicago.edu}
         \and
             Departamento de F\'isica e Astronomia, Faculdade de Ci\^encias,
             Universidade do Porto, Rua do Campo Alegre, 4169-007 Porto,
             Portugal
             }
   \date{Received \ldots,\ldots; accepted \ldots,\ldots}
  \abstract
{The Rossiter-McLaughlin (hereafter RM) effect is a key tool for
measuring the projected spin-orbit angle between stellar spin axes and
orbits of transiting planets. However, the measured radial velocity (RV)
anomalies produced by this effect are not intrinsic and depend on both
instrumental resolution and data reduction routines. Using inappropriate
formulas to model the RM effect introduces biases, at least in the
projected velocity $V\sin i_\star$ compared to the spectroscopic value.
Currently, only the iodine cell technique has been modeled, which
corresponds to observations done by, e.g., the HIRES spectrograph of the
Keck telescope. In this paper, we provide a simple expression of the RM
effect specially designed to model observations done by the Gaussian fit
of a cross-correlation function (CCF) as in the routines performed by
the HARPS team. We derived also a new analytical formulation of the RV
anomaly associated to the iodine cell technique. For both formulas, we
modeled the subplanet mean velocity $v_p$ and dispersion $\beta_p$
accurately taking the rotational broadening on the subplanet profile
into account. We compare our formulas adapted to the CCF technique with
simulated data generated with the numerical software SOAP-T and find
good agreement up to $V\sin i_\star\lesssim 20$ km.s$^{-1}$. In
contrast, the analytical models simulating the two different observation
techniques can disagree by about 10$\sigma$ in $V\sin i_\star$ for large
spin-orbit misalignments. It is thus important to apply the adapted
model when fitting data.
}
   \maketitle
\section{Introduction}
Transiting planets produce radial velocity (RV) anomalies when crossing the
disk of their star. This mechanism, known as the Rossiter-McLaughlin
effect (hereafter RM effect), is due to the stellar proper rotation and
the fact that during a transit, a planet successively covers different
portions of the stellar disk with different average velocities along the
line of sight \citep{Holt_AA_1893, Rossiter_ApJ_1924,
McLaughlin_ApJ_1924}. The RM effect has gained importance in the exoplanet
community since it allows the measurement of the projected angle between
the stellar spin-axis and the orbit of the planet.

The first measurements of the RM effect induced by a transiting planet
were performed almost simultaneously with two different instruments on
the same bright star HD209458.  \citet{Queloz_etal_AA_2000} observed the
signal with the ELODIE spectrograph on the 193cm telescope of the
Observatoire de Haute Provence, while \citet{Bundy_Marcy_PASP_2000} used
the HIRES spectrograph on the Keck telescope. As explained later on, the
choice of the instrument and, more particularly, the subsequent
reduction analysis, have a non negligible impact on the resulting shape
of signal. It is thus interesting to see that the two kinds of
instrument, coupled with their own data treatment, which are still
employed today, have been used in parallel since the beginning.

The first planet-host stars studied with this technique were all
compatible with low obliquity. It was along the lines of the model of
planet migration in a protoplanetary disk. But then, several misaligned
systems have been detected, starting with XO-3 with an angle initially
announced at $70^\circ\pm15^\circ$ \citep{Hebrard_etal_AA_2008} and then
refined to $37.3^\circ\pm3^\circ$ \citep{Winn_etal_ApJ_2009}, yet still
significantly misaligned. With the growth of the sample, a first
correlation appeared showing that the hottest stars tend to be more
misaligned \citep{Winn_etal_ApJ_2010}.  Additionally, a first
statistical comparison between observations and theoretical predictions
have been performed \citep{Triaud_etal_AA_2010}, suggesting that some of
the hot Jupiters might be the result of the interaction with a stellar
companion, leading to a Lidov-Kozai mechanism, characterized by phases of
large eccentricity and inclination and followed by a circularization by
tides raised on the planet as it approaches the star
\citep{Wu_Murray_ApJ_2003, Fabrycky_Tremaine_ApJ_2007}. Other scenarios
have then been developed to explain the formation of hot Jupiters, such
as planet-planet scattering \citep{Rasio_Ford_Science_1996,
Beauge_Nesvorny_ApJ_2012}, the crossing of secular resonances
\citep{Wu_Lithwick_ApJ_2011}, or the Lidov-Kozai mechanism produced by a
planetary companion \citep{Naoz_etal_Nature_2011,
Nagasawa_Ida_ApJ_2011}. A new trend between age and obliquity has also
been found, suggesting that tidal dissipation may play an important role
in the evolution of those systems \citep{Triaud_AA_2011}.

Accurate modelings of the RM effect are thus needed to get reliable
information on the current obliquity of stars and to test theoretical
predictions. In the literature, one can find several analytical
expressions to model this effect \citep{Kopal_PNAS_1942,
Ohta_etal_ApJ_2005, Gimenez_ApJ_2006, Hirano_etal_ApJ_2010,
Hirano_etal_ApJ_2011}. They are not all identical because they model
different techniques of radial velocity measurements. 

This raises an issue that should be considered with caution.  RVs
measured by different techniques or even by different instruments using
the same algorithm can differ by more than a constant offset. To
illustrate this point, we consider a random variable with a given
probability distribution function (PDF), and ask what its {\em average}
value is. The term {\em average} is vague and can mean different
quantities: mean, median, mode, etc. Nevertheless, if the PDF is
symmetrical, any of these quantities lead to the same result, eventually
with different robustnesses with respect to noise.  But if the PDF is
{\em not} symmetrical, each estimator of the {\em average} value
provides different results that cannot be compared directly. The
situation is similar in RV observations, as noticed by
\citet{Hirano_etal_ApJ_2010}. There are at least two different ways to
measure RVs. One relies on the iodine cell technique which consists in
fitting an observed spectrum with a modeled one that is Doppler-shifted
\citep{Butler_etal_PASP_1996}.  The other is based on a Gaussian fit to
a cross-correlation function (CCF) \citep{Baranne_etal_AA_1996,
Pepe_etal_AA_2002}.  The former technique is applied to observations
made with HIRES on the Keck telescope or with HDS at the Subaru
telescope, while the latter is the routine of, e.g., SOPHIE at the
Observatoire de Haute Provence or HARPS at La Silla Observatory.
 If stars are affected by neither spots nor transiting planets, their
spectral lines have constant shapes, and thus the RVs derived by any
instruments may only differ by a constant offset. In contrast, spectral
lines are deformed during transits, and these deformations vary with
time. As a consequence, each analysis routine is expected to lead to a
different signal. In turn, it is important to have an analytical model
adapted to each analysis routine in order to interpret RM data.
Moreover, it also means that one should not combine RM measurements from
different instruments.

The first analytical expressions of the RM effect were derived by
\citet{Kopal_PNAS_1942}, \citet{Ohta_etal_ApJ_2005}, and
\citet{Gimenez_ApJ_2006}. They all computed the weighted mean
velocity, hereafter called $\va$, along the line of sight of the
stellar surface uncovered by the planet. This mean velocity,
weighted by the surface intensity, leads to an exact expression of the
form
\be
\va = -\frac{f}{1-f}v_p\ ,
\label{eq.ohta}
\ee
where $f$ is the fraction of the flux blocked by the planet disk and
$v_p$ is the average velocity of the surface of the star covered by the
planet. This expression is simple and exact. There is no assumption
behind it. But, it does not correspond to the quantity that is actually
measured in the observations by either the iodine cell technique, or
by the Gaussian fit of the CCF.  Thus, the analytical prediction $\va$
is systematically biased when compared directly with observations. The
difference increases for high stellar rotational velocities.

\figexpl

To solve this problem, \citet{Hirano_etal_ApJ_2010} propose a new
analytical expression closely related to the reduction algorithm
of the iodine cell technique (see Fig.~\ref{fig.explanation}).
This method consists in fitting a shifted spectrum outside of transits,
which is modeled, here, by a single averaged spectral line and
noted\footnote{Here, and throughout the paper, the line profiles are
expressed as a function of radial velocity $v$, instead of wavelength.}
$\Fstar(v-\vx)$, with the line profile detected during a transit
$\Ftrans(v)$, via the Doppler shift $\vx$.  The quantity provided by the
formulas of \citet{Hirano_etal_ApJ_2010} is the value, hereafter denoted
$\vb$, of the velocity $\vx$ that maximizes the cross-correlation
between the two line profiles $C(\vx) = \int \Fstar(v-\vx) {\cal F}_{\rm
transit}(v)\,dv$.  We show in Sect.~\ref{sec.app_Keck} that if $\Fstar$
is an even function, the maximization of the cross-correlation $C(\vx)$
is indeed identical to the minimization of the chi-square associated to
the fit of $\Ftrans(v)$ by $\Fstar(v-\vx)$. In that case, the result
depends on the actual shape of the line profiles, as observed in
practice.  In the simplest case where the line profiles of both the
nonrotating and the rotating star are Gaussian, this method leads to
\citep{Hirano_etal_ApJ_2010}
\be
\vb = -\left(\frac{2\beta_\star^2}{\beta_\star^2+\beta_p^2}\right)^{3/2}
f v_p \left(1-\frac{v_p^2}{2(\beta_\star^2+\beta_p^2)}\right)\ ,
\label{eq.hirano}
\ee
where $\beta_\star$ and $\beta_p$ are the dispersion of the Gaussian
profiles $\Fstar(v)$ and $\Fpla(v-v_p)$ respectively\footnote{In this paper,
unit Gaussians are defined by $\Gfit(v)=(\sqrt{2\pi}\sigma)^{-1} 
\exp{(-v^2/(2\sigma^2))}$, while in \citep{Hirano_etal_ApJ_2010}, they are defined by
${\cal G}_{\tilde \sigma}(v)=(\sqrt{\pi}\tilde\sigma)^{-1}
\exp{(-v^2/\tilde\sigma^2)}$. There is thus a difference of a
factor 2 in the parenthesis of Eq.~(\ref{eq.hirano}) with respect to 
\citep[][Eq.~36]{Hirano_etal_ApJ_2010}.}. $\Fpla(v-v_p)$ is
the line profile of the light blocked by the planet centered
on $v_p$. Although this expression is the result of an expansion in
both $f$ and $v_p$, it gives a better representation of the measured
radial velocity. \citet{Hirano_etal_ApJ_2010} also provide more complex
expressions in the case of Voigt profiles, and in
\citet{Hirano_etal_ApJ_2011}, additional effects are taken into account
such as macro-turbulence. But the result is not expressed as a
simple analytical formula and it requires several numerical integrations.

The model of \citet{Hirano_etal_ApJ_2010,Hirano_etal_ApJ_2011} is well
adapted to the iodine cell technique, but it still does not correspond
to the analysis routines used on stabilized spectrographs, e.g.,
SOPHIE and HARPS \citep{Baranne_etal_AA_1996, Pepe_etal_AA_2002}. In
these routines, the observed spectrum is first cross-correlated with a
template spectrum.  This gives the so-called cross-correlation function
(CCF) which can be seen as a weighted average of all the spectral lines
convolved with a rectangular function.  Finally, the CCF is fitted by a
Gaussian whose mean represents the observed radial velocity $\vc$
(see Fig.~\ref{fig.explanation}).  Currently, there is no
analytical expression of this quantity in the literature.  The goal of
this paper is to provide such an expression.  As we will see, even in
the general case where the spectral line profile $\Fstar$ is not
Gaussian, the resulting formula is as simple as Eq.  (\ref{eq.hirano})
derived by \citet{Hirano_etal_ApJ_2010} and exact in
$v_p$.

This paper is organized as follows. In Sect.~\ref{sec.gen_expression},
we analytically derive an unbiased expression $\vc$ modeling the RM
effect as measured by, e.g., SOPHIE and HARPS. This formula is specially
designed to simulate the radial velocity measurements obtained by fitting a
Gaussian to a CCF. We first provide very generic expressions that relies
only on the symmetry of the spectroscopic lines, and then, we give a much
simpler formula corresponding to Gaussian subplanet line profiles. In
Sect.~\ref{sec.iodine}, we propose a new expression $\vd$ of the RM
signal derived by the iodine technique. In comparison to the previous
ones, it is analytical and valid for any spectroscopic $V\sin
i_\star$. Then, in Sect.~\ref{sec.fvpbetap}, we detail the calculation
of the parameters entering into our formulas $\vc$ and $\vd$, which are the
flux fraction $f$ occulted by the planet, the subplanet velocity $v_p$,
and the dispersion $\beta_p$. In Sect.~\ref{sec.comparison}, we compare
our analytical results with simulated data generated with SOAP-T,
a modified version of the numerical code SOAP
\citep{Boisse_etal_AA_2012}, able to reproduce RM signals
\citep{Oshagh_etal_AA_2012}. We also analyze biases introduced by
the application of a wrong model in the fit of RM signals. Finally, we
conclude in Sect.~\ref{sec.conclusion}.

\section{Modeling of the RM effect measured by the CCF technique}
\label{sec.gen_expression}

\subsection{General derivation}
We derive a very general expression of the RM effect obtained using
a Gaussian fit to the CCF. As in \citet{Hirano_etal_ApJ_2010}, we
only consider the linear effect with respect to the flux ratio $f$. But
the method can easily be generalized to higher orders. We define
$\Fstar(v)$, $\Fpla(v-v_p)$, and $\Ftrans(v)$ the line profile of the
CCF produced by the integrated stellar surface, by the part of the
stellar surface covered by the planet, and by the uncovered stellar
surface, respectively. At this stage, the line profiles
$\Fstar$, $\Fpla$, and $\Ftrans$ are not necessarily Gaussian. The
dependency of $\Fpla$ is on $(v-v_p)$ because this line is centered on
$v_p$.  By convention, $\Fstar(v)$ and $\Fpla(v-v_p)$ are normalized to
one. With this convention, even absorption spectral lines are {\em
positive}. We have $\Ftrans(v) = \Fstar(v)
- f\Fpla(v-v_p)$.  Moreover, we denote ${\cal G}_{\sigma}(v)$ as the
  unit Gaussian profile with dispersion $\sigma$ and centered on the
origin
\be
{\cal G}_{\sigma}(v) = \frac{1}{\sqrt{2\pi}\sigma} 
\exp\left({-\frac{v^2}{2\sigma^2}}\right)\ .
\ee
The fit of the CCF measured during transit, $\Ftrans(v)$, by a Gaussian
corresponds to the maximization of the likelihood
\be
\chi^2(a,\vx,\sigma) = \int\big(\Ftrans(v)-a\Gfit(v-\vx)\big)^2\,dv
\ee
with respect to the normalization factor $a$, the mean velocity $\vx$, and
the dispersion $\sigma$. The partial derivatives read as
\be
\EQM{
\Dron{\chi^2}{a} &= -2 \int \Gfit(v-\vx) \big(
\Ftrans(v)-a\Gfit(v-\vx)\big)\,dv \ ; \crm
\Dron{\chi^2}{\vx} &= -\frac{2a}{\sigma^2} \int (v-\vx) \Gfit(v-\vx)
\big( \Ftrans(v)-a\Gfit(v-\vx)\big)\,dv \ ; \crm
\Dron{\chi^2}{\sigma} &= -\frac{2a}{\sigma^3} \int \!((v-\vx)^2\!-\!\sigma^2) \Gfit(v\!-\!\vx)
\big( \Ftrans(v)-a\Gfit(v\!-\!\vx)\big)\,dv .
}
\label{eq.sys1}
\ee
We now set all the derivatives to zero, express $\Ftrans(v)$ as a function of
$\Fstar(v)$ and $\Fpla(v)$, and reorder the terms. The system of
equations (\ref{eq.sys1}) is equivalent to
\be
\EQM{
\int     F(a,\vx,\sigma; v)\,dv &= f\int     \Fpla(v-v_p)\Gfit(v-\vx)\,dv \ ;\crm
\int v   F(a,\vx,\sigma; v)\,dv &= f\int v   \Fpla(v-v_p)\Gfit(v-\vx)\,dv \ ;\crm
\int v^2 F(a,\vx,\sigma; v)\,dv &= f\int v^2 \Fpla(v-v_p)\Gfit(v-\vx)\,dv \ ,
}
\label{eq.sys2}
\ee
with
\be
F(a,\vx,\sigma; v) = \Gfit(v-\vx)\big(\Fstar(v)-a\Gfit(v-\vx)\big)\ .
\ee

To solve this system, we apply the usual perturbation method. We develop
the parameters in series of the flux ratio $f$, i.e., $a=a_0 + f a_1 +
f^2 a_2 + \ldots$, and idem for $\vx$ and $\sigma$. At the zeroth order
in $f$, the system (\ref{eq.sys2}) corresponds to the fit of $\Fstar(v)$
by a Gaussian. The effect of the planet is absent from the fit, thus
$\vx_0=0$. The other parameters $a_0$ and $\sigma_0$ are those of the
best Gaussian fit outside of transits.

In a second step, we linearize the system (\ref{eq.sys2}) in the
vicinity of the zeroth order solution. We thus compute the partial
derivatives of 
\be
M_n(a,\vx,\sigma) = \int v^n F(a,\vx,\sigma)\, dv\ ,
\ee
appearing in the left-hand side of (\ref{eq.sys2}),
at $a=a_0$, $\vx=\vx_0=0$, and $\sigma=\sigma_0$. To avoid fastidious
calculation, we make the simple hypothesis that $\Fstar(v)$ is
symmetrical, or more precisely, an even function of the velocity
$v$. In practice, due to the convective blue shift (CB), spectral
lines are not perfectly symmetric. Nevertheless, the net effect of CB is
to add a constant offset that does not modify RM signals
\citep{Albrecht_etal_ApJ_2012}.
Then, $M_0(a,\vx,\sigma)$ and $M_2(a,\vx,\sigma)$ are even 
functions of $\vx$, while $M_1(a,\vx,\sigma)$ is odd. As a result, the
derivatives ${\partial M_0}/\partial \vx$, $\partial M_2/\partial \vx$,
$\partial M_1/\partial a$, and $\partial M_1/\partial \sigma$ taken at
$\vx=0$ vanish. Moreover, if the subplanet line profile is also an even
function then the integrals of the righthand side of the system
(\ref{eq.sys2}) become simple convolutions at $\vx=0$. 
Then there is only
\be
\EQM{
 \left(\Dron{M_0}{a}\right) a_1 
+\left(\Dron{M_0}{\sigma}\right) \sigma_1 &= 
\left[\Gfitb*\Fpla\right](v_p)\ , \crm
 \left(\Dron{M_1}{\vx}\right) \vx_1 &= 
\left[(v\,\Gfitb)*\Fpla\right](v_p)\ , \crm
 \left(\Dron{M_2}{a}\right) a_1 
+\left(\Dron{M_2}{\sigma}\right) \sigma_1 &= 
\left[(v^2\Gfitb)*\Fpla\right](v_p)\ ,
}
\ee
where $*$ denotes the convolution product.
The first and the third lines are independent of the mean velocity
$\vx_1$. Their resolution provides a correction to the amplitude and the
width of the best Gaussian fit during a transit. The second equation is
the most interesting, since it contains the quantity $\vx_1$ we are looking 
for. By chance, this is also the simplest. The velocity anomaly obtained 
by the Gaussian fit $\vc = \vx = f \vx_1$ is then
\be
\vc = f\left(\Dron{M_1}{\vx}\right)^{-1}
\left[(v\,\Gfitb)*\Fpla)\right](v_p)
\ee
where $\partial M_1/\partial \vx$ should be computed at $(a_0, 0,
\sigma_0)$, i.e.,
\be
\EQM{
\Dron{M_1}{\vx} &= 
\int \frac{v^2}{\sigma_0^2}\Gfitb(v)
\left(\Fstar(v)-2a_0\Gfitb(v)\right)\,dv
\crm &=
-\frac{a_0}{4\sigma_0\sqrt{\pi}} + 
\left[(v^2\Gfitb)*(\Fstar-a_0\Gfitb)\right](0)\ .
}
\label{eq.dM1dv}
\ee
The convolution product taken at 0 on the righthand side of
(\ref{eq.dM1dv}) is nothing else but $M_2$, which cancels by the
definitions of $a_0$ and $\sigma_0$. Thus the RM effect now reads as
\be
\vc = -\frac{4\sigma_0\sqrt{\pi}}{a_0} f
\left[(v\,\Gfitb)*\Fpla)\right](v_p)\ .
\label{eq.solgen}
\ee
This expression does not depend directly on the stellar line profile
$\Fstar$. The dependence only occurs through the best Gaussian fit
($a_0$ and $\sigma_0$), which is performed to derive the radial velocity.
The formula (\ref{eq.solgen}) is thus very powerful, since it does not
require any knowledge on $\Fstar$. Unfortunately, the amplitude $a_0$ of
the best Gaussian fit in (\ref{eq.solgen}) is associated to a normalized
line profile $\Fstar$ while, in practice, the area of a CCF is difficult
to measure, and $\Fstar$ is never normalized. We thus provide the
expression of $a_0$ as a function of $\Fstar$ and the best Gaussian fit
$\Gfitb$,
\be
a_0 = \frac{\int \Gfitb(v)\Fstar(v)\,dv}{\int\Gfitb(v)\Gfitb(v)\,dv}
    = 2\sigma_0\sqrt{\pi} \left[\Gfitb*\Fstar\right](0)\ .
\label{eq.a0}
\ee
We emphasize that $a_0$ is independent of $v_p$. As a
consequence, it does not affect the shape of the RM effect, but only
slightly the amplitude ($a_0$ remains close to one). The
computation of $a_0$ is detailed in Appendix~\ref{sec.app_a0}.

\subsection{Gaussian subplanet line profile}
At this stage, the expression (\ref{eq.solgen}) is very general, and
holds as long as the line profiles $\Fstar(v)$ and $\Fpla(v)$ are
symmetric. 

We now make the hypothesis that the subplanet line profile $\Fpla(v)$ is
Gaussian. It should be stressed that, as long as the planet radius is
small compared to that of the star, the subplanet line profile is only
weakly affected by the stellar rotation (see Sect.~\ref{sec.fvpbetap})
and thus, it is well approximated by that of the nonrotating star, which
we assume to be Gaussian. Then, the subplanet profile can be considered
Gaussian, or equal to the sum of two Gaussians if macro-turbulence is
taken into account (see Appendix~\ref{sec.appMT}).  We denote $\beta_p$
as the width of the subplanet line profile, i.e., $\Fpla(v)={\cal
G}_{\beta_p}(v)$.  In that case, the expression of the RM effect
(\ref{eq.solgen}) becomes
\be
\vc = -\frac{1}{a_0}
\left(\frac{2\sigma_0^2}{\sigma_0^2+\beta_p^2}\right)^{3/2}
f v_p \exp\left(-\frac{v_p^2}{2(\sigma_0^2+\beta_p^2)}\right)\ .
\label{eq.solexact}
\ee
This is the main equation of this paper. It represents a good
compromise between simplicity and accuracy for the modeling of RM 
signals measured by a Gaussian fit of the CCF.

\subsection{Gaussian stellar line profile}
\label{sec.gaussapprox}
For completeness, we give the expression in the case where the stellar
line profile is a normalized Gaussian with dispersion $\beta_\star$.
The best fit should give $a_0=1$ and $\sigma_0=\beta_\star$. Then, we
get
\be
\vc = -\left(\frac{2\beta_\star^2}{\beta_\star^2+\beta_p^2}\right)^{3/2}
f v_p \exp\left(-\frac{v_p^2}{2(\beta_\star^2+\beta_p^2)}\right)\ .
\label{eq.solgauss}
\ee
This formula is equivalent to $\vb$ (\ref{eq.hirano}) given by
\citet{Hirano_etal_ApJ_2010}. More precisely, Eq.~(\ref{eq.hirano}) is 
the beginning of the expansion of $\vc$. Indeed, if the stellar line
profile is Gaussian, the two approaches are identical.

\section{Modeling of the RM effect measured by the iodine cell
technique}
\label{sec.iodine}
In this section, we first explain the equivalence between the iodine
cell technique and the maximization of the cross-correlation $C(\vx)$
\citep{Hirano_etal_ApJ_2010, Hirano_etal_ApJ_2011}. The aim is to
emphasize the hypotheses behind this equivalence and, thus, to show its
limitations. In a second step, we provide a general expression that models
the iodine cell technique. It should be stressed that the function
$C(\vx)$ is different from the CCF of the previous section and is not
used in the same way. It involves the spectrum during transit and
a modeled one without transit deformations. This function $C(\vx)$ is
computed to provide the RV at its maximum. On the other hand,
the CCF is the cross-correlation between the spectrum and a mask.  The
goal is to provide a single averaged line that is then fitted by a
Gaussian curve.
 
\subsection{Link between the iodine cell technique and the maximization
of $C(\vx)$}
\label{sec.app_Keck}
The analysis routine based on the iodine cell technique involves a fit
with 13 parameters of the observed spectrum by a modeled one that is
Doppler-shifted \citep{Butler_etal_PASP_1996}. We assume that this 
can be approximated by the fit of a single parameter ($\vx$)
representing the Doppler shift between a modeled line profile $\Fstar(v)$
and the observed one (here during a transit) $\Ftrans(v)$. The
chi-square of this fit reads
\be
\chi^2(\vx) = \int_{-\infty}^{+\infty} \big(\Ftrans(v)-\Fstar(v-\vx)\big)^2\,dv\ .
\ee
The minimization of this chi-square corresponds to $d
\chi^2/d \vx = 0$ with
\be
\frac{d\chi^2(\vx)}{d\vx} = 2\int
\frac{d\Fstar}{dv}(v-\vx)\big(\Ftrans(v)-\Fstar(v-\vx)\big)\,dv\ .
\ee
The integral on the righthand side can be split into the sum of two
integrals
\be
\frac{d\chi^2(\vx)}{d\vx} = 2\big(I_1(\vx)+I_2(\vx)\big)\ ,
\ee
where the first one, $I_1(\vx)$, is the cross-correlation of 
$d\Fstar/d v$ by $\Ftrans$ taken at $\vx$, while the other is,
after the change of variable $u=v-\vx$,
\be
I_2 = \int_{-\infty}^{\infty} \frac{d\Fstar}{dv}(u) \Fstar(u)\,du\ .
\ee
If $\Fstar$ is even, its derivative is odd, and thus the integral $I_2$ 
over $\mathbb{R}$ vanishes. In that case, only
\be
\frac{d\chi^2(\vx)}{d\vx} = 2\left(\frac{d\Fstar}{dv}\right)
   \star \Ftrans
\ee
remains, where $\star$ denotes the cross-correlation product defined by
\be
[f\star g](x) = \int_{-\infty}^{+\infty} f(y-x)g(y)\,dy\ .
\ee
We then use the property of the derivative of the cross-correlation
of two functions
\be
\frac{d}{dx}(f\star g) = -\frac{df}{dx}\star g = f\star\frac{dg}{dx}\ .
\ee
By identification, we obtain
\be
\frac{d\chi^2(\vx)}{d\vx} = -2\frac{d}{d\vx} (\Fstar\star \Ftrans) 
= -2\frac{dC(\vx)}{d\vx}\ ,
\ee
where $C(\vx)$ is defined as in \citet{Hirano_etal_ApJ_2010}. Thus, the
minimization of the chi-square involved in the iodine cell technique is
indeed equivalent to the maximization of the cross-correlation $C(\vx)$ 
as computed in \citet{Hirano_etal_ApJ_2010}. This result holds as long
as the complicated fit with 13 parameters can be modeled by the fit of the
single parameter $\vx$, and if the modeled line profile is symmetrical.
The second condition may not be true in general. If the asymmetry is not
too strong, the integral $I_2$ would be a small perturbation, and the
result obtained by the maximization of the cross-correlation $C(\vx)$
should differ from the minimization of the chi-square by only a small
constant.

\subsection{General expression of the RM effect}
\label{sec.gen_iodine}
To derive a general expression of the RV signal measured by the iodine
technique, we use the same model as \citet{Hirano_etal_ApJ_2010}, which
consists in maximizing $C(\vx)$ where
\be
C(\vx) = \int \Fstar(v-\vx)
\big[\Fstar(v)-f\Fpla(v-v_p)\big]\,dv\ .
\ee
Then, the condition $dC(\vx)/d\vx=0$ leads to
\be
\left[\frac{d\Fstar}{dv}\star \Fstar\right](\vx) = 
f \int \frac{d\Fstar}{dv}(v-\vx) \Fpla(v-v_p)\,dv\ .
\ee
As in the previous section, we expand $\vx$ in series of $f$:
$\vx = \vx_0 + f\vx_1 + ...$. At the zeroth order, we get
\be
\left[\frac{d\Fstar}{dv}\star \Fstar\right](\vx_0) = 0\ .
\ee
If $\Fstar(v)$ is an even function, this equality gives $\vx_0=0$.
Otherwise, $\vx_0$ would be a small constant, depending only on the
shape of the spectral lines, but not on $v_p$. Here, we assume that
$\vx_0=0$. At the first order in the flux ratio $f$, assuming that 
$\Fpla$ is even, we obtain, with $\vd=f\vx_1$,
\be
\vd = \frac{1}{A_0}f\left[\frac{d\Fstar}{dv} * \Fpla\right](v_p)\ ,
\label{eq.vdconv}
\ee
where
\be
A_0 = \frac{d}{d\vx}\left[\frac{d\Fstar}{dv}\star\Fstar\right]_{\vx=0}
    = \int_{-\infty}^{+\infty} \left(\frac{d\Fstar}{dv}\right)^2\,dv\ .
\ee
This result is more complex than (\ref{eq.solgen}) because it involves
the derivatives of the stellar line profile $\Fstar$ instead of the
derivatives of a best Gaussian fit $\Gfitb$ which are analytical.
Of course, if $\Fstar$ is Gaussian, we retrieve the result of the
previous section (see Sect.~\ref{sec.gaussapprox}).

In Appendix~\ref{sec.appdGconvR}, we provide an analytical
expression of the numerator of $\vd$ (\ref{eq.vdconv}) in the case 
where the subplanet line profile is Gaussian and for a general
limb-darkening law. 

\section{Parameters of the subplanet line profile}
\label{sec.fvpbetap}
The expressions of the RM effect (\ref{eq.solexact}) and
(\ref{eq.vdconv}) depend on the fraction $f$ of the flux covered by the
planet, the subplanet velocity $v_p$, and the dispersion $\beta_p$. There
are two approaches to evaluate them. On
the one hand, both the flux fraction $f$ and the mean velocity $v_p$ can
be computed exactly as a series of Jacobi polynomials
\citep{Gimenez_ApJ_2006}. This is useful in the case of binary transits
where the occulting object is big. On the other hand, only the flux
fraction $f$ is derived exactly using analytical algorithms such as the
one given by \citep{Mandel_Agol_ApJ_2002}, while $v_p$ and $\beta_p$ are
estimated assuming uniform intensity below the planet
\citep{Hirano_etal_ApJ_2010, Hirano_etal_ApJ_2011}. Then, if $(\bar x,
\bar y)$ are the averaged coordinates over the surface of the star
covered by the planet, and normalized to the radius of the star,
$v_p\approx\bar{x} V\sin i_\star$, while $\beta_p$ is constant and
represents the width $\beta_0$ of the nonrotating star line profile.

Here, we choose a compromise between the two approaches and take the
slope and the curvature of the intensity below the planet into account.
This gives a better estimate of $v_p$ and $\beta_p$ in comparison to the
uniform subplanet intensity hypothesis. But also it turns out that the
method provides a simple and accurate expression for the flux fraction
$f$.  Another advantage of this method is that it can be easily applied
to more complex problems where the gravity-darkening or the tidal
deformations of both the planet and the star are taken into account.

\subsection{Method}
To describe the method, we take the example of the computation of the
flux fraction $f$. The expression of $f$ reads as
\be
f = \iint_\Spla I(x,y)\,dx\,dy\ ,
\ee
where $\Spla$ is the surface of the stellar disk covered by the planet
normalized by the square of the radius of the star.
$I(x,y)$ is the normalized limb-darkening of the star expressed as a function
of the normalized coordinates $(x,y)$. For the moment, we do not need to
give its expression.

A very rough approximation of $f$ is obtained assuming uniform intensity
below the planet. In that case, we obtain
\be
\EQM{
f &= I(\bar x, \bar y) \iint_\Spla \,dx\,dy 
\crm &= I(\bar x, \bar y) \,\Spla\ ,
}
\label{eq.frough}
\ee
where
\be
\bar x = \frac{1}{\Spla}{\iint_\Spla x\, dx\, dy}\ , 
\quad \textrm{and}\quad
\bar y = \frac{1}{\Spla}{\iint_\Spla y\, dx\, dy}\ ,
\label{eq.xybar}
\ee
are the coordinates of the barycenter of the portion of the stellar disk
covered by the planet.  The formula (\ref{eq.frough}) works well for
very small planets but only during full transits
\citep[e.g.][]{Mandel_Agol_ApJ_2002}. Close to the limb, the intensity
$I(x,y)$ varies strongly with position and this approximation is not
valid anymore. To overcome this issue, we propose to make an expansion
of the limb-darkening profile $I(x,y)$ in the vicinity of $(\bar x, \bar
y)$. We get
\be
\EQM{
\frac{f}{\Spla} 
  &= I(\bar x, \bar y) \moy{}{1}
   + J^{(0)}_x \moy{}{x-\bar x}
+J^{(0)}_y \moy{}{y-\bar y}
+\frac{1}{2}H^{(0)}_{xx} \moy{}{(x-\bar x)^2}
\crm &
+\frac{1}{2}H^{(0)}_{yy} \moy{}{(y-\bar y)^2}
+H^{(0)}_{xy} \moy{}{(x-\bar x)(y-\bar y)}\ ,
}
\label{eq.ff}
\ee
where for any function $f(x,y)$, 
\be
\moy{}{f(x,y)} = \frac{1}{\Spla}\iint_\Spla f(x,y)\,dx\,dy \ ,
\label{eq.moyf}
\ee
where $J^{(0)}_x = \partial_x I(x,y)$ and $J^{(0)}_y=\partial_y I(x,y)$ 
are the components of the Jacobian of the surface
intensity $I(x,y)$ computed at $(\bar x, \bar y)$. 
Similarly, the components of the Hessian are
\be
H^{(0)}_{xx} = \Dron{^2I(x,y)}{x^2}\ ,\quad
H^{(0)}_{yy} = \Dron{^2I(x,y)}{y^2}\ ,\quad
H^{(0)}_{xy} = \Dron{^2I(x,y)}{x\partial y}\ .
\ee
By construction, $\bar x$ and $\bar y$ are defined by $\bar
x=\moy{}{x}$, and $\bar y=\moy{}{y}$. Thus, the linear terms in factor of
the Jacobian $(J^{(0)}_x, J^{(0)}_y)$ cancel in (\ref{eq.ff}). At that
point, only
\be
\EQM{
\frac{f}{\Spla} =& I(\bar x, \bar y)
+\frac{1}{2}H^{(0)}_{xx} \moy{}{(x-\bar x)^2}
+\frac{1}{2}H^{(0)}_{yy} \moy{}{(y-\bar y)^2}
\crm &
+ H^{(0)}_{xy} \moy{}{(x-\bar x)(y-\bar y)} 
}
\label{eq.fsimp}
\ee
remains.
The first term in (\ref{eq.fsimp}) corresponds to the rough
approximation derived in Eq.~(\ref{eq.frough}). The other terms provide a
correction proportional to the square of the normalized planet radius
($r=R_p/R_\star$) and are expected to be small.

\subsection{Subplanet velocity}
The above method applied to $I(x,y)$ to get the flux fraction $f$ can be
adapted to any other function. For example, the subplanet velocity is
defined by
\be
v_p = V\sin i_\star \frac{\iint_\Spla x I(x,y)\,dx\,dy}
{\iint_\Spla I(x,y)\, dx\,dy}\ .
\label{eq.vp0}
\ee
We denote $H^{(1)}_{xx}$, $H^{(1)}_{yy}$, and $H^{(1)}_{xy}$ as the
components of the Hessian of $xI(x,y)$ at the averaged position 
$(\bar x, \bar y)$ (\ref{eq.xybar}). Using the expression of $f$ (\ref{eq.fsimp}), 
at first order in $r^2$, we get
\be
\EQM{
\frac{v_p}{V\sin i_\star} =& \bar x 
+ \frac{1}{2I(\bar x, \bar y)}\Bigg(\left(H^{(1)}_{xx}-\bar xH^{(0)}_{xx}\right)
\moy{}{(x-\bar x)^2}
\crm &
+ \left(H^{(1)}_{yy}-\bar xH^{(0)}_{yy}\right) \moy{}{(y-\bar y)^2}
\crm &
+ 2\left(H^{(1)}_{xy}-\bar xH^{(0)}_{xy}\right) \moy{}{(x-\bar x)(y-\bar y)}
\Bigg)\ .
}
\label{eq.vp}
\ee
The denominator $I(\bar x, \bar y)$ in (\ref{eq.vp}), as
well as the terms in $H^{(0)}_{xx}$, $H^{(0)}_{yy}$, and $H^{(0)}_{xy}$,
come from the expansion of the denominator of (\ref{eq.vp0}).

\subsection{Width of the subplanet line profile}
The width of the subplanet line profile $\beta_p$ is a combination of the
width of the nonrotating line profile $\beta_0$ and a correction
$\delta \beta_p$ due to the rotational broadening
\be
\beta_p^2 = \beta_0^2 + \delta\beta_p^2\ .
\ee
We define ${\cal V}_2$ as the average of the square of the subplanet
velocity
\be
{\cal V}_2 = \left(V\sin i_\star\right)^2
\frac{\iint_\Spla x^2I(x,y)\,dx\,dy}{\iint_\Spla I(x,y)\,dx\,dy}\ .
\label{eq.V2}
\ee
With this notation, we have
\be
\delta\beta_p^2
= {\cal V}_2
- v_p^2\ .
\label{eq.betap}
\ee
Noting $H_{xx}^{(2)}$, $H_{yy}^{(2)}$, and $H_{xy}^{(2)}$, the components
of the Hessian of $x^2I(x,y)$ at the averaged coordinates $(\bar x, \bar
y)$, the expression of ${\cal V}_2$ reads as
\be
\EQM{
\frac{{\cal V}_2}{(V\sin i_\star)^2}
=& {\bar x}^2 +\frac{1}{2I(\bar x, \bar y)}\Bigg(
\left(H^{(2)}_{xx}-\bar x^2H^{(0)}_{xx}\right)\moy{}{(x-\bar x)^2}
\crm &
+ \left(H^{(2)}_{yy}-\bar x^2H^{(0)}_{yy}\right) \moy{}{(y-\bar y)^2}
\crm &
+ 2\left(H^{(2)}_{xy}-\bar x^2H^{(0)}_{xy}\right) \moy{}{(x-\bar x)(y-\bar y)}
\Bigg)\ .
}
\ee
At first order, the ${\bar x}^2$ in ${\cal V}_2$ (\ref{eq.V2}) cancels 
with the square of $\bar x$ in the expression of $v_p$ (\ref{eq.vp}).
Thus, at first order, $\delta\beta_p$ vanishes and the width of the subplanet
profile is equal to the width of the nonrotating star $\beta_0$.
However, the quadratic terms do not cancel, and this provides an
estimation of the contribution of the rotational broadening to the
actual width of the subplanet profile.

\subsection{Limb-darkening and its derivatives}
As we saw above, the determination of the subplanet profile depends on
the limb-darkening law and its second derivatives. In this section,
we provide generic formulas assuming that the limb-darkening law is a
linear combination of functions $I_\alpha(x,y)$ defined by
\be
I_\alpha(x,y) = \mu^\alpha = \left(1-x^2-y^2\right)^{\alpha/2}\ ,
\label{eq.Ialpha}
\ee
where $\mu=\sqrt{1-x^2-y^2}$ is the cosine of the angle between the
normal of the stellar surface at $(x,y)$ and the observer.

The second derivatives $H_{xx}^{(n)}$, $H_{yy}^{(n)}$, and $H_{xy}^{(n)}$
of $x^nI_\alpha(x,y)$ are given Table~\ref{tab.Hessian}. These are the
ones that are needed to compute the flux fraction $f$, the subplanet
velocity $v_p$ and the dispersion $\beta_p$. In practice, only the cases
$n=0,1,2$ are used.

\tabHess

From these general formulas, one can derive the expressions for the
quadratic limb-darkening which reads as
\be
\EQM{
I_q(x,y) &= I_q(0)\left(1-u_1(1-\mu)-u_2(1-\mu)^2\right)
\crm &= u'_0 + u'_1\mu + u'_2\mu^2\ ,
}
\ee
with $I_q(0)$ the central intensity such that $I_q(x,y)$ is
normalized to one. By identification, we get
\be
\EQM{
u'_0 &= \frac{1-u_1-u_2}{\pi(1-u_1/3-u_2/6)}\ , \crm 
u'_1 &= \frac{u_1+2u_2}{\pi(1-u_1/3-u_2/6)}\ , \crm
u'_2 &= \frac{-u_2}{\pi(1-u_1/3-u_2/6)}\ .
}
\label{eq.uprime}
\ee
Equivalently, the so-called nonlinear limb-darkening is usually
expressed as
\be
\EQM{
I_{nl}(x,y) &= I_{nl}(0) \left(1-\sum_{n=1}^4 c_n \left(1-\mu^{n/2}\right)\right)
\crm &
= c'_0 +c'_1\mu^{1/2}+c'_2\mu+c'_3\mu^{3/2}+c'_4\mu^{2}\ ,
}
\ee
with
\be
c'_0 = \frac{1-c_1-c_2-c_3-c_4}{\pi(1-c_1/5-c_2/3-3c_3/7-c_4/2)}\ ,
\label{eq.c0prime}
\ee
and for $1\leq n\leq4$,
\be
c'_n = \frac{c_n}{\pi(1-c_1/5-c_2/3-3c_3/7-c_4/2)}\ .
\label{eq.cnprime}
\ee
The normalizations in (\ref{eq.uprime}), (\ref{eq.c0prime}),
and (\ref{eq.cnprime}) have been deduced from the integral of each
$I_\alpha(x,y)$ over the entire disk of the star
\be
\iint \mu^\alpha\,dx\,dy 
= \int_0^{2\pi}d\phi\int_0^{\pi/2}d\theta\,
\cos^{\alpha+1}\theta\sin\theta
= \frac{2\pi}{\alpha+2}\ .
\ee

\subsection{Averaged coordinates and covariances}
The last quantities that need to be computed in order to get the
subplanet line profiles are the averaged coordinates $\moy{}{x}$,
$\moy{}{y}$, the variances $\moy{}{(x-\bar x)^2}$, 
$\moy{}{(y-\bar y)^2}$, and the covariance 
$\moy{}{(x-\bar x)(y-\bar y)}$. For that, we distinguish two cases.

\subsubsection{During a full transit}
In the case of a full transit, i.e., when the disk of the planet is fully
inside the disk of the star, the problem gets simpler since the
integrals (\ref{eq.moyf}) have to be computed over a uniform disk of
area $\Spla=\pi r^2$ and centered on the coordinates $(x_0,y_0)$ of the
planet. We get $(\bar x, \bar y) = (x_0, y_0)$, and
\be
\EQM{
\moy{}{(x-\bar x)^2} = \moy{}{(y-\bar y)^2} = \frac{r^2}{4}\ ,
\crm 
\moy{}{(x-\bar x)(y-\bar y)} = 0\ .
}
\ee

\subsubsection{During ingress or egress}
If the disk of the planet is crossing the limb of the star, the area
where the integrals of the form (\ref{eq.moyf}) are computed is not
circular (see the shaded area in Fig.~\ref{fig.coord}). In that case, we
use the very powerful method of \citet{Pal_MNRAS_2012}, which gives
expressions that also work also for mutual transits.
\figCoord

We recall briefly the method that relies on Green's theorem
converting an integral over a surface into an integral over the contour
of that surface:
\be
\iint_{\cal S} \omega(x,y)\,dx\,dy = 
\oint_{\partial \cal S} \left(f_x(x,y)\,dx+f_y(x,y)\,dy\right)\ .
\label{eq.green}
\ee
In this equation, $\partial {\cal S}$ is the boundary of ${\cal S}$, and
$\omega(x,y)$ the exterior derivative of $\vec f(x,y) = (f_x,f_y)$
defined by
\be
\omega(x,y) = \Dron{f_y}{x} - \Dron{f_x}{y}\ .
\ee
When the planet is crossing the limb of the star, the boundary is the
union of two circular arcs. One of them follows the edge of the planet
centered on $(x_0,y_0)$ with radius $r$. The coordinates of any points of
this arc and the tangent vectors are of the form
\be
\EQM{
x &= x_0 + r\cos\varphi\ , \crm
y &= y_0 + r\sin\varphi\ ,
}
\qquad
\EQM{
dx &= -r\sin\varphi\,d\varphi\ , \crm
dy &=  r\cos\varphi\,d\varphi\ .
}
\ee
The angle $\varphi$ varies between two limits $\varphi_p^{(1)}$ and 
$\varphi_p^{(2)}$, corresponding to the intersections $A$ and $B$ 
between the circumferences of the planet and of the star, respectively
(see Fig.~\ref{fig.coord}). The second arc fellows the edge of the star
and is parameterized by the coordinates
\be
\EQM{
x &= \cos\varphi\ , \crm
y &= \sin\varphi\ ,
}
\qquad
\EQM{
dx &= -\sin\varphi\,d\varphi\ , \crm
dy &=  \cos\varphi\,d\varphi\ , 
}
\ee
with $\varphi$ going from $\varphi_s^{(1)}$ to $\varphi_s^{(2)}$
associated to the intersections $B$ and $A$, respectively.

More generally, if we denote $(x_j, y_j)_{j=p,s}$ as the centers of the arcs,
and $(r_j)_{j=p,s}$ as their radii such that 
\be
x(\varphi)=x_j+r_j\cos\varphi\ , \qquad{\rm and}\qquad
y(\varphi)=y_j+r_j\sin\varphi\ ,
\ee
we obtain \citep{Pal_MNRAS_2012}
\be
\EQM{
&\iint_\Spla \omega(x,y)\,dx\,dy = 
\crm &
\sum_{j=p,s}
\int_{\varphi_j^{(1)}}^{\varphi_j^{(2)}} \left(
 f_y(x(\varphi),y(\varphi))\cos\varphi
-f_x(x(\varphi),y(\varphi))\sin\varphi\right)\,r_j\,d\varphi\ .
}
\ee
Here, we are interesting in the cases where $\omega(x,y)$ stands 
for $1$, $x$, $y$, $x^2$, $y^2$, or $xy$. The field 
vectors $\vec f(x,y)$ associated to those $\omega(x,y)$ are not 
uniquely determined. We choose $(-\frac{1}{2}y, \frac{1}{2}x)$,
$(-xy,0)$, $(0,xy)$, $(-x^2y,0)$, $(0,xy^2)$, and
$(-\frac{1}{4}xy^2,\frac{1}{4}x^2y)$, respectively. We denote
$\omega_{ij}=x^iy^j$ and ${\vec f}_{ij}(x,y)$ as the functions whose 
exterior derivative is $\omega_{ij}$. The integral of 
${\vec f}_{ij}$ along a circular arc reads as
\be
\EQM{
F_{ij}(\varphi) = \int &\big( 
  f_{ij,y}(x_0+r\cos\varphi,y_0+r\sin\varphi)\cos\varphi
\crm &
-f_{ij,x}(x_0+r\cos\varphi,y_0+r\sin\varphi)\sin\varphi\big)
\,r\,d\varphi\ .
}
\label{eq.Fij}
\ee
Since the ${\vec f}_{ij}$ are polynomials in $x$ and $y$, the $F_{ij}$
can be computed using the recurrence relations provided by
\citet{Pal_MNRAS_2012}. The results are displayed in
Table~\ref{tab.integFij} for $i=0,1,2$, and $j=0,1,2$. From them, one can
derive the quantities present in the expressions of the flux fraction
$f$, the subplanet velocity $v_p$ and the width $\beta_p$:
\tabintegFij
\be
\EQM{
\Spla = \moy{}{1} = \iint \omega_{00}\ , \crm 
\bar x = \moy{}{x} = \frac{1}{\Spla} \iint \omega_{10}\ , \crm
\bar y = \moy{}{y} = \frac{1}{\Spla} \iint \omega_{01}\ , \crm
\moy{}{(x-\bar x)^2} = \left(\frac{1}{\Spla} \iint \omega_{20}\right) 
 - \bar x^2 \ , \crm
\moy{}{(y-\bar y)^2} = \left(\frac{1}{\Spla} \iint \omega_{02}\right) 
 - \bar y^2 \ , \crm
\moy{}{(x-\bar x)(y-\bar y)} = 
  \left(\frac{1}{\Spla} \iint \omega_{11}\right)
- \bar x \bar y\ .
}
\ee

\section{Comparison with simulations}
\label{sec.comparison}

\subsection{Transit light curve}
Although it was not the main goal of this present work, in the
derivation of a precise modeling of the RM effect, we obtained a new
expression of the flux fraction $f$ occulted by a planet during a
transit (see Eq.~(\ref{eq.fsimp})). In comparison to existing formulas
that are exact \citep[e.g.][]{Mandel_Agol_ApJ_2002, Pal_MNRAS_2012},
the one of this paper relies on an expansion of the intensity in the
vicinity of the averaged position of the planet. We thus expect our
formulation to be less precise.
\figflux

Figure \ref{fig.flux} shows the comparison between the approximation
(\ref{eq.fsimp}) and the exact formula derived by
\citet{Mandel_Agol_ApJ_2002}. By eye, it is not possible to distinguish
between the two approaches. In the residuals, however, we can see that
the maximum of deviation occurs close to the limb, more exactly, when
the edge of the planet is tangent to that of the star. Indeed, at the border of the star,
the limb-darkening becomes steeper and steeper, and the derivatives
$\partial_x I_\alpha(x,y)$ and $\partial_y I_\alpha(x,y)$ even go to
infinity for $\alpha<2$. Nevertheless, this singularity is smoothed out
by the decrease in the overlapping area between the planet and the star
disks during ingress and egress. 

One advantage of the present formula is that it can be easily
generalized to more complex problems, as in the cases of a distorted
planet, distorted star, important gravity limb-darkening, and so on. For
our purpose, it provides an accurate enough estimation of the flux that
can be used to derive the RM effect.

\subsection{Subplanet profile}
\label{sec.profile}
We checked the accuracy of our new formulas of the subplanet
velocity $v_p$ (\ref{eq.vp}) and the width $\beta_p^2 = \beta_0^2 +
\delta\beta_p^2$ with $\delta\beta_p$ given by (\ref{eq.betap}).
For that, we used the software called SOAP, for Spot Oscillation And
Planet \citep{Boisse_etal_AA_2012}, to produce artificial data as close
as possible to real observations. This code is a numerical tool that
models radial velocity and photometry observations of stars with spots.
It has been updated recently to also model the effect of a planet
transiting a spotted star, and was renamed SOAP-T
\citep{Oshagh_etal_AA_2012}. Briefly, the code
divides the disk of the star into a grid. To each cell of that grid, a
Gaussian profile with a width $\beta_0$ and amplitude $I(x,y)$ (in our
notation) is assigned. This represents the intrinsic line profile of the
nonrotating star as detected by the instrument. These lines are then
shifted in velocity according to their position with respect to the
spin-axis and the $V\sin i_\star$ of the star. All the lines of the
cells uncovered by any spots or planets are added together to produce
an artificial CCF that is then fitted by a Gaussian to derive a radial
velocity.

With SOAP-T, we produced the CCF of a star with a transiting
planet at different positions of the planet on the disk. We also
generated the CCF of the same star while the planet is not transiting,
and by taking the difference, we got the subplanet profile. Such profiles are
displayed in Fig.~\ref{fig.planetprofile} for different values of $V\sin
i_\star$.  Unless specified explicitly, here, and in all the following
simulations, the star is a solar-type star with a quadratic
limb-darkening law whose coefficients are $u_1=0.38$, $u_2=0.3$, and an
intrinsic line width without rotation of $\beta_0=3$ km.s$^{-1}$. The
planet is a Jupiter evolving in the equatorial plane of its star, its
radius is $r=R_p/R_{\rm star} = 0.1099$.  In
Fig.~\ref{fig.planetprofile}, the subplanet line profiles of low
rotating stars are Gaussian. This results from the hypothesis of
SOAP-T, which assumes Gaussian intrinsic line profiles. But we
observe that the Gaussian shape holds even for $V\sin i_\star=20$
km.s$^{-1}$, which validates our assumption leading to
Eq.~(\ref{eq.solexact}).
\figplanetprofile

To each of the artificial subplanet profiles generated with
SOAP-T, we also computed the mean velocity $v_p$ and the
dispersion $\beta_p$, to be compared with our formulas (\ref{eq.vp}) and
(\ref{eq.betap}). Figure \ref{fig.vp} shows the results for $v_p$ after
normalization to remove the effect of the $V\sin i_\star$ of the star.
We checked that the figure is indeed unchanged up to $V\sin i_\star=20$
km.s$^{-1}$.  The numerical outputs obtained with SOAP-T are
plotted against two different analytical approximations denoted $S_0$ and
$S_2$. In $S_0$, the surface brightness of the star is taken uniform
below the disk of the planet, while in $S_2$, the second derivatives are
taken into account as in (\ref{eq.vp}). We observe that where the error
is maximal, close to the limb, $S_2$ improves the determination of $v_p$
by about a factor 3 with respect to $S_0$. In the case $r=0.1$ and
$V\sin i_\star=10$ km.s$^{-1}$, the maximal error provided by $S_2$ is
about 20 m/s which represents a relative difference of 0.2\%.
\figvp

In the case of the dispersion $\beta_p$, the difference between the
estimation derived assuming uniform ($S_0$) and nonuniform ($S_2$)
brightness below the planet disk is more evident (see
Fig.~\ref{fig.betap}). Indeed, in the former case, $\beta_p$ remains
constant and equal to the width $\beta_0$ of the nonrotating star line
profile, while we observe that for the simulated and the modeled line
profiles, the shape of $\beta_p$ as a function of the orbital phase
looks like a trapezoid with the large base at $\beta_0$ and the maximum
at approximately $\sqrt{\beta_0^2+(r\, V\sin i_\star)^2/4}$.
\figbetap

\subsection{Rossiter-McLaughlin effect}
We now compare our analytical expression of the Rossiter-McLaughlin
effect $\vc$ (\ref{eq.solexact}) with signals generated with
SOAP-T, which simulates the reduction analysis of the CCF
technique numerically.

\figRM

Figure~\ref{fig.RM} displays the results for different $V\sin i_\star$.
As long as $V\sin i_\star$ is below or equal to 10 km.s$^{-1}$, the error
induced by the analytical formula remains lower than $\sim$1 m/s, which
is close to the magnitude of the precision of RV measurements. In
that case, the analytical approximations are almost indistinguishable
from the numerical simulations. However, for larger $V\sin i_\star$, the
agreement between numerical signals and analytical ones is weaker. For
example, when $V\sin i_\star=20$ km.s$^{-1}$, the analytical approximation
leads to a maximal error of 10 m/s, which is 5\% of the amplitude of the
signal. Nevertheless, it should also be noted that for fast-rotating stars
the spreading of the spectral lines over the detectors decreases the
precision of the measurements. 
In any case, the analytical expression $\vc$ brings a definite improvement
over other formulas, which have not been designed to simulate
the CCF technique as we see in the following section.

\subsection{Comparison between different techniques}
\label{sec.comp}
\figLines
\figRMcomp
To highlight the effect of the instrument and of the data
reduction analysis, we generated different models of line profile and
compared the RM signals computed numerically with the results of the
analytical formulas $\vc$ (\ref{eq.solexact}) and $\vd$
(\ref{eq.vdconv}).  In our examples, the line profiles are of three
types: ${\cal F}_{\rm HIRES}(v)$, ${\cal F}_{\rm HARPS}(v)$, and ${\cal
F}_{\rm CORALIE}(v)$. They are associated to three RM signals: $v_{\rm
HIRES}$, $v_{\rm HARPS}$, and $v_{\rm CORALIE}$, respectively.  It
should be stressed that the goal is not to reproduce the lines
observed by those instruments exactly, but to capture their main characteristics.
On the one hand, HIRES and HARPS are two spectrographs with high
resolutions that we assume to be identical with a width
$\beta_0=2.6$ km.s$^{-1}$ for nonrotating solar-type stars. On the
other hand, the resolution of CORALIE is about twice lower, and the
intrinsic width of the same stars is about $\beta_0=4.5$ km.s$^{-1}$
\citep{Santos_etal_AA_2002}. We consider a star with $V\sin i_\star=15$
km.s$^{-1}$, which is adapted to our illustration.  Finally, the
transiting planet is a Jupiter-like planet with a radius $R_p=0.1R_{\rm
star}$. 

Figure~\ref{fig.linetest} shows the simulated line profiles. The
panel \ref{fig.linetest}a displays the rotation kernel ${\cal R}(v)$,
a stellar profile $\Fstar(v)$ with $\beta_0=2.6$ km.s$^{-1}$, and a
subplanet profile $\Fpla(v-v_p)$ multiplied by the flux fraction $f$,
and computed with $\beta_p = (\beta_0^2+(r V\sin i_\star/4)^2)^{1/2}$.
Figure~\ref{fig.linetest}b depicts the resulting line profiles
during transit ${\cal F}_{\rm HIRES}(v)$, ${\cal F}_{\rm HARPS}(v)$, and
${\cal F}_{\rm CORALIE}(v)$. Following our hypothesis, ${\cal F}_{\rm
HIRES}(v)$ is identical to ${\cal F}_{\rm HARPS}(v)$.

From the simulated line profiles, we derived RM signals numerically.
The signal $v_{\rm HIRES}$ was obtained from ${\cal F}_{HIRES}$ using the iodine
cell technique, i.e., by fitting the best Doppler shift between a line
without transit deformation, and the line profiles computed during
transit. Both $v_{\rm HARPS}$, and $v_{\rm CORALIE}$ are the results of
applying the CCF technique, i.e., a numerical fit between a shifted
Gaussian and ${\cal F}_{\rm HARPS}$ and ${\cal F}_{\rm CORALIE}$,
respectively. We also generated $v_{{\cal C}\rm max}$ by maximizing the
cross-correlation $C(\vx)$ between the line profiles ${\cal F}_{\rm
HIRES}$ at and out of transit.  These four RM signals are represented in
Fig.~\ref{fig.RMcomp}a. It is notable that the RM effects associated
to the three instruments are all different. The variation between
$v_{\rm HARPS}$ and $v_{\rm CORALIE}$ is only due to the change in
resolution. However, in the case of $v_{\rm HIRES}$ and $v_{\rm HARPS}$,
the simulated lines are exactly identical. The observed difference in
the RM signal is the result of the chosen data reduction technique.
Figure~\ref{fig.RMcomp}a also confirms that the maximum of the
cross-correlation $C(\vx)$ gives the same result as the iodine cell
technique (when the stellar lines are symmetrical) since $v_{\rm HIRES}
= v_{{\cal C}\rm max}$.

The last three panels of Fig.~\ref{fig.RMcomp} represent the
comparison between the simulated RM signals and the analytical formulas
$\vc$ (\ref{eq.solexact}), and $\vd$ (\ref{eq.vdconv}) associated to the
CCF and the iodine cell technique, respectively. We observe that the
formulas adapted to the analysis routines are in good agreement with the
respective simulations. We also notice that for CORALIE, whose
resolution is lower, the two analytical formulas give roughly the same
result. This is because the stellar line is less affected
by the rotational kernel and is more Gaussian.  We show that, in that
case, the two methods should indeed provide the same result (see
Sect.~\ref{sec.gaussapprox}).

From this study, we conclude that a given star observed by two different
techniques should present two distinct RM signals. To date, this
notable result has not been seen since the instruments with the highest
signal-to-noise, HIRES and HARPS, are located in two different hemispheres.
This makes it difficult to observe the same stars. For those
observed with other instruments, the expected gaps are diluted
by the measurement uncertainties. Nevertheless, with the advent of
HARPS-North, we may observe such discrepancies in the future.

\subsection{Biases on fitted parameters}
\label{sec.bias}

As a final test, we simulated artificial data from either the CCF or the
iodine cell model, and we fit each of these data with the two models
separately. The goal is not to perform an exhaustive study of the biases
introduced by the application of a wrong model in the process of fitting 
data, but to give an example with some typical parameters.

For this illustration, we considered only one set of parameters. As in the
previous section, the star has $V \sin i_\star=15$ km.s$^{-1}$ with
intrinsic line width $\beta_0=2.6$ km.s$^{-1}$. We chose a quadratic
limb-darkening characterized by $u_1=0.69$ and $u_2=0.0$. The planet's
radius is taken equal to $R_p = 0.1 R_{\rm star}$.  The impact parameter
of the orbit is assumed to be $0.3 R_{\rm star}$.  For information, this
value is that of a planet with a semi-major axis $a = 4 R_{\rm star}$
and an inclination $i = 85.7$ deg. All these parameters were fixed
throughout all the simulations. Only the projected spin-orbit angle
$\lambda_{\rm input}$ was varied from 0 to 90 degrees by steps of ten
degrees. For each value of $\lambda_{\rm input}$, and each model, 1\,000
datasets were generated with a Gaussian noise of 10 m/s.  Each simulation
contains 50 points, among which 32 are inside the transit and 18
outside. In each case, we fit both the $V\sin i_\star$ and the projected
spin-orbit angle.

Figure~\ref{fig.bias_CCF} shows the results of this analysis in the case
where the data are generated with the CCF model $\vc$
(Eq.~(\ref{eq.solexact})). As expected, the parameters recovered with
the appropriate model are accurate, while those deduced from the iodine
cell technique formulas are biased. The bias on $V \sin i_\star$ is
systematically positive and also the most important, especially at large
projected spin-orbit angle ($\lambda_{\rm input}\approx90$ deg) where we
get $(V\sin i_\star)_{\rm fit}=20.7\pm0.5$ km.s$^{-1}$ instead of 15
km.s$^{-1}$. One can notice that this agrees with the results
of, e.g., \citet{Simpson_etal_MNRAS_2010} who applied the model of
\citeauthor{Hirano_etal_ApJ_2010} on WASP-3 observed with SOPHIE. They
fit a $V\sin i_\star=15.7^{+1.4}_{-1.3}$ km.s$^{-1}$ while the
spectroscopic value is only $13.4\pm1.5$ km.s$^{-1}$. On the other hand,
the bias on the fitted projected spin-orbit angle $\lambda_{\rm fit}$
remains within 2-$\sigma$. This parameter is thus less affected by the
model.

The difference in behavior between $(V \sin i_\star)_{\rm fit}$ and
$\lambda_{\rm fit}$ is more evident in Fig.~\ref{fig.bias_iodine}. In
that case, the data were simulated with the formulas associated to the
iodine cell technique: $\vd$ (Eq.~(\ref{eq.vdconv})). As in the previous
test, using the same model for both the generation of the data and the
fit, leads to very accurate estimations of the parameters, while the
application of the wrong model introduces biases. The $(V \sin
i_\star)_{\rm fit}$ is systematically negative as we could expect since
the situation is the opposite of the one in the previous paragraph.
Nevertheless, the error is smaller. In the worst case, we get $(V \sin
i_\star)_{\rm fit} = 11.9\pm0.3$ km.s$^{-1}$, which represents on error
of the order of 3 km.s$^{-1}$, while it was almost 6 km.s$^{-1}$ in the
previous example. The situation is similar for $\lambda_{\rm fit}$. We
observe small biases anticorrelated with those of the previous test, but
now, the inaccuracy remains within 1-$\sigma$.

We stress that we only fit two parameters in this study, while the others
are fixed to their exact values. We already observe that the best fits
tend to compensate for inaccurate models by introducing biases. With more
free parameters, there are more possibilities to balance the model, and 
it is thus difficult to predict the behavior of the fit. Since the
models are not linear, we should expect the presence of several local 
minima. Eventually, in some of them, the projected spin-orbit angle
might be more biased than in our tests. This should be analyzed on
individual case bases, which is not the goal of this paper.

\figBiasa
\figBiasb

\section{Conclusion}
\label{sec.conclusion}

One of the main objectives of this paper has been to highlight that
there is no unique way of measuring RM effects and that different
techniques provide different values of RV anomalies. RM signals should
thus be analyzed using the appropriate model to avoid any biases, at
least in $V\sin i_\star$. This is particularly important in the case of 
low-impact parameters (planet passing close to the center of its star)
since then, the projected spin-orbit angle only depends on the
amplitude of the RM signal.

We provided a new analytical formula specially designed to
model RV anomalies obtained by fitting a Gaussian function to the CCF,
as in the analysis routines of HARPS and SOPHIE. We also revisited the
modeling of the iodine cell technique, as used with HDS and HIRES, for
which we derived an analytical expression adapted to non-Gaussian stellar
line profiles. An effort was made to model the effect of the
rotation of the star on the width of the subplanet line profile. Since
our formulas do not rely on any expansion in powers of the subplanet
velocity $v_p$, our results remain stable even for fast-rotating stars.

The advantage of having a purely analytical expression to model the RM
effect is the rapidity of computation. It can thus be used to analyze
a large sample of RM signals uniformly.
As a complement to this paper, we make our code accessible to the
community as a free open source software package. This is a library 
called ARoME, an acronym for Analytical Rossiter-McLaughlin Effect,
designed to generate analytical RM signals based on the formulas of the
paper. It also includes the effect of macro-turbulence as described in
the Appendix~\ref{sec.appMT}. The library provides a C interface and,
optionally, a Fortran 77 or 2003 interface to be called by an
application. The fully documented package can be downloaded from the
webpage \url{www.astro.up.pt/resources/arome}.

Besides the modeling of the RM effect, we also analytically derived a
new expression for transit light curves (\ref{eq.fsimp}). Although this
expression is the result of a Taylor expansion of the intensity and is
only adapted to small planets, it gives good approximations.  Moreover,
the expression is general enough to be easily extended to more
complex problems.

\section*{Acknowledgments}
We thank Amaury Triaud for helpful discussions and feedback on this
project. We also acknowledge the support by the European Research
Council/European Community under the FP7 through Starting Grant
agreement number 239953, as well as from Funda\c{c}\~ao para a Ci\^encia
e a Tecnologia (FCT) in the form of grant reference
PTDC/CTE-AST/098528/2008.  NCS also acknowledges the support from FCT
through program Ci\^encia\,2007 funded by FCT/MCTES (Portugal) and
POPH/FSE (EC). MM and IB would furthermore like to thank the FCT for fellowships
SFRH/BPD/71230/2010 and SFRH/BPD/81084/2011, respectively.

\appendix

\section{Macro-turbulence}
\label{sec.appMT}
Here, we study the effect of macro-turbulence on the Rossiter-McLaughlin
signal. We consider only the ``radial-tangential'' model as in
\citep{Hirano_etal_ApJ_2011}. In that case, if we denote ${\cal F}_{\rm
0}(v)$ as the line profile of the nonrotating star without
macro-turbulence, the subplanet line profile reads as
\be
\Fpla(v) = \left[{\cal F}_{\rm 0} * M\right](v)\ ,
\label{eq.FplaM}
\ee
where $M(v)$ is the rotation-turbulence kernel given by 
\citep{Gray_book_2005},
\be
M(v) = \iint_{\Spla} I(x,y) \Theta(x,y,v-x V\sin i_\star)\, dx\,dy\ ,
\ee
and
\be
\EQM{
\Theta(x,y,v) &= 
\frac{1}{2}\big(\Theta_R(x,y,v)+\Theta_T(x,y,v)\big)
\crm & =
\frac{1}{2\sqrt{\pi}}\left(
 \frac{1}{\zeta\cos\theta} \e^{-\left(\frac{v}{\zeta\cos\theta}\right)^2}
+\frac{1}{\zeta\sin\theta} \e^{-\left(\frac{v}{\zeta\sin\theta}\right)^2}
\right)\ .
}
\label{eq.Theta}
\ee
We highlight the different dependencies on $(x,y)$, on the one hand,
through $\cos\theta=\sqrt{1-x^2-y^2}$ and $\sin \theta=\sqrt{x^2+y^2}$, 
and on the velocity $v$, on the other. The coordinates $(x,y)$ are
normalized by the radius of the star.
Since $\Theta$ is the sum of two Gaussians $\Theta_R$ and $\Theta_T$
associated to the radial and the tangential broadenings, respectively,
we also split $M(v)$ into two parts
\be
M(v) = \frac{1}{2}\big(M_R(v)+M_T(v)\big)\ ,
\ee
such that $M_R(v)$ is associated to $\Theta_R$, and 
$M_T(v)$ is associated to $\Theta_T$.

Now, we compute the moments of $(M_j(v))_{j=R,T}$ as in
Sect.~\ref{sec.fvpbetap} to evaluate the effect of the
rotation-turbulence kernel on the subplanet profile. We have
\be
\EQM{
\moy{M_j}{v^n} = \frac{1}{A_j} & \int_{-\infty}^{+\infty} dv\, v^n
\crm & \times
\iint_{\Spla} dx\,dy\, I(x,y) \Theta_j(x,y,v-xV\sin i_\star)\ ,
}
\label{eq.vn}
\ee
where $A_j$ is a normalization constant whose expression is
\be
A_j = \int_{-\infty}^{+\infty} dv \iint_{\Spla}
dx\,dy\, I(x,y) \Theta_j(x,y,v-xV\sin i_\star)\ .
\ee
Inverting the integrals, we get, for the normalization,
\be
A_j = \iint_{\Spla} dx\,dy\,I(x,y) \int_{-\infty}^{+\infty}
dv\,\Theta_j(x,y,v-xV\sin i_\star)\ .
\ee
Since $\Theta_j(x,y,v)$ (\ref{eq.Theta}) is normalized, the inner
integral on the velocity is one. It thus remains only
\be
A_R = A_T = A = \iint_{\Spla} I(x,y)\,dx\,dy\ ,
\ee
as in Sect.~\ref{sec.fvpbetap}. We now focus on the numerator of 
(\ref{eq.vn}). By construction, $\moy{M_j}{v^0}=1$. Then, using the
inversion of integrals, we get
\be
\EQM{
v_p^{(j)} := \moy{M_j}{v} = \frac{1}{A} \iint_{\Spla} &dx\,dy\,\bigg[I(x,y) 
\crm & \times
\int_{-\infty}^{+\infty} dv\,v\, \Theta_j(x,y,v-xV\sin i_\star)\bigg]\ .
}
\ee
The inner integral over the velocity $v$ gives $xV\sin i_\star$, we have
thus
\be
v_p^{(j)} = V\sin i_\star \frac{\iint_{\Spla} xI(x,y)\,dx\,dy}
{\iint_{\Spla} I(x,y)\,dx\,dy}
\ee
for each broadening: radial ($j=R$) and tangential ($j=T$). This is identical
to (\ref{eq.vp0}).  Finally, the second moment reads as
\be
\EQM{
{\cal V}_2^{(j)} := \moy{M_j}{v^2} = \frac{1}{A} &\iint_{\Spla} dx\,dy\,
\bigg[ I(x,y)
\crm & \times
\int_{-\infty}^{+\infty} dv\,v^2\Theta_j(x,y,v-xV\sin i_\star)\ .
}
\ee
The inner integral gives
\be
\int_{-\infty}^{+\infty} dv\,v^2\Theta_R(v-xV\sin i_\star)
 = \zeta^2\cos^2\theta+x^2 (V\sin i_\star)^2
\ee
for the radial broadening, and
\be
\int_{-\infty}^{+\infty} dv\,v^2\Theta_T(v-xV\sin i_\star)
 = \zeta^2\sin^2\theta+x^2 (V \sin i_\star)^2
\ee
for the tangential broadening. We thus have
\be
{\cal V}_2^{(R)} = \zeta_R^2 + {\cal V}_2\, \quad
{\rm and}\quad
{\cal V}_2^{(T)} = \zeta_T^2 + {\cal V}_2\ ,
\ee
with
\be
\zeta_R^2 = \zeta^2
\frac{\iint_{\Spla} \cos^2\theta \,I(x,y)\,dx\,dy}
{\iint_\Spla I(x,y)\,dx\,dy}\ ,
\ee
\be
\zeta_T^2 = \zeta^2
\frac{\iint_{\Spla} \sin^2\theta \,I(x,y)\,dx\,dy}
{\iint_\Spla I(x,y)\,dx\,dy}\ , 
\ee
where ${\cal V}_2$ corresponds to the case without macro-turbulence 
(Eq.~(\ref{eq.V2})). It should be noted that $\zeta_R^2+\zeta_T^2=\zeta^2$.
Let $\beta_0$ be the dispersion of ${\cal F}_{\rm 0}(v)$, and $\delta
\beta_p$ the dispersion due to the rotational broadening alone
(\ref{eq.betap}). The subplanet line profile $\Fpla$ (\ref{eq.FplaM})
can be approximated by the sum of two Gaussian functions
\be
\Fpla(v-v_p) = \frac{1}{2}\big({\cal G}_{\beta_R}(v-v_p)+{\cal G}_{\beta_T}(v-v_p)\big)
\ee
centered on the same value $v_p$ with respective dispersions
\be
\beta_R^2 = \beta_0^2+\delta\beta_p^2+\zeta_R^2\ ,
\ee
and
\be
\beta_T^2 = \beta_0^2+\delta\beta_p^2+\zeta_T^2\ .
\ee
In this expressions, $\delta\beta_p$, $\zeta_R$, and $\zeta_T$ are
functions of the position of the planet on the stellar disk.
With this model, the Rossiter-McLaughlin effect, as measured by the
Gaussian fit of the CCF, reads as
\be
\EQM{
\vc =& -\frac{1}{2a_0} \left(\frac{2\sigma_0^2}{\sigma_0^2+\beta_R^2}\right)^{3/2}
f v_p \exp\left(-\frac{v_p^2}{2(\sigma_0^2+\beta_R^2)}\right)
\crm &
       -\frac{1}{2a_0} \left(\frac{2\sigma_0^2}{\sigma_0^2+\beta_T^2}\right)^{3/2}
f v_p \exp\left(-\frac{v_p^2}{2(\sigma_0^2+\beta_T^2)}\right)\ .
}
\ee

\section{Normalization factor of the Gaussian fit}
\label{sec.app_a0}

\subsection{Without macro-turbulence}
\label{sec.a0wout}
Here we detail the computation of the amplitude $a_0$ of the best
Gaussian fit (\ref{eq.a0}). In a first step we neglect the
macro-turbulence and have
\be
a_0 = 2\sigma_0\sqrt{\pi}\left[\Gfitb * \Fstar \right](0)\ ,
\ee
with $\Fstar = {\cal F}_{\rm 0} * {\cal R}$ and ${\cal R}$ is the
normalized rotation kernel (\ref{eq.Rv}). If we assume that ${\cal
F}_{\rm 0}$ is Gaussian with dispersion $\beta_0$, the associativity of
the convolution product leads to
\be
\EQM{
a_0 &= 2\sigma_0\sqrt{\pi}\left[{\cal G}_{\sigma_t} * {\cal R} \right](0)
\crm &
 = 2\sigma_0 \sqrt{\pi} \int_{-V\sin i_\star}^{V\sin i_\star} 
{\cal G}_{\sigma_t}(v) {\cal R}(v) \,dv\ ,
}
\ee
with $\sigma_t^2=\sigma_0^2+\beta_0^2$. Finally, since ${\cal
G}_{\sigma_t}(v)$ and ${\cal R}(v)$ are even functions of $v$, the
expression of $a_0$ can be slightly simplified
\be
a_0 = 4\sigma_0 \sqrt{\pi} \int_0^{V\sin i_\star} 
{\cal G}_{\sigma_t}(v) {\cal R}(v) \,dv\ .
\label{eq.appa0}
\ee
The amplitude $a_0$ is thus given by one single integral over a finite
interval, which only has to be computed once. It can be done numerically
using, for example, the simple trapezoidal rule explained in
\citet{Press_etal_book_1992}. In the case of quadratic limb-darkening,
the result can also be expressed explicitly as a combination of
modified Bessel functions and error functions
\citep[e.g.][Eq.~F5]{Hirano_etal_ApJ_2010}.

\subsection{With macro-turbulence}
If the macro-turbulence is taken into account, it is not anymore possible 
to express the amplitude $a_0$ as a simple integral as in
Sect.~\ref{sec.a0wout}. This is because the
rotational-macroturbulence broadening kernel cannot be expressed as a
convolution product \citep{Gray_book_2005}. But since $a_0$ brings only
a small correction with respect to the Gaussian case ($a_0$ remains
close to 1), we simplify the problem and approximate the line profile of
the nonrotating star as a single Gaussian with dispersion $\beta'_0$ given by
\be
\beta_0^{\prime 2} = \beta_0^2+\frac{\zeta^2}{2}
\ee
instead of two Gaussians with dispersion $\beta_{0,R}$ and
$\beta_{0,T}$ defined by
\be
\beta_{0,R}^2 = \beta_0^2 + \zeta^2\cos^2\theta
\quad {\rm and}\quad
\beta_{0,T}^2 = \beta_0^2 + \zeta^2\sin^2\theta\ .
\ee
With this simplification, we recover the expression (\ref{eq.appa0})
where $\sigma_t$ as to be replaced by $\sigma'_t$ defined by
$\sigma_t^{\prime 2} = \sigma_0^2 + \beta_0^{\prime 2}$, i.e.,
\be
a_0 = 4\sigma_0 \sqrt{\pi} \int_0^{V\sin i_\star} 
{\cal G}_{\sigma'_t}(v) {\cal R}(v) \,dv\ .
\label{eq.appa0'}
\ee

\subsection{Rotation kernel}
We now give the expression of the rotation kernel ${\cal R}(v)$ present
in the expression of the amplitude $a_0$ (\ref{eq.appa0}) and
(\ref{eq.appa0'}), without or with macro-turbulence, respectively.
First, we consider a simpler kernel ${\cal R}_\alpha(v)$ associated to
an intensity $I_\alpha(x,y)$ of the form
\be
I_\alpha(x,y) = (1-x^2-y^2)^{\alpha/2}\ .
\ee
With $u=v/(V\sin i_\star)$, the rotation kernel \citep{Gray_book_2005} reads as
\be
{\cal R}_\alpha(v) = \frac{1}{V\sin i_\star}
\int_{-\sqrt{1-u^2}}^{\sqrt{1-u^2}} (1-u^2-y^2)^{\alpha/2}\,dy\ .
\ee
To simplify the integral, we make the change of variable
$z=y/\sqrt{1-u^2}$. We obtain
\be
{\cal R}_\alpha(v) = \frac{b(\alpha)}{V\sin i_\star}
\left(1-\left(\frac{v}{V\sin i_\star}\right)^2\right)^{(\alpha+1)/2}\ ,
\ee
where
\be
b(\alpha) = \int_{-1}^1 (1-z^2)^{\alpha/2}\,dz\ .
\ee
In practice, for the quadratic and the nonlinear limb-darkening, only
the cases $\alpha=n/2$, $n\in\mathbb{N}$ are used. The integrals
$b(\alpha)$ can thus be computed using the recurrence relation
\be
b(\alpha) = \frac{\alpha}{\alpha+1}b(\alpha-2)
\ee
with the initial conditions
\be
\EQM{
b(-1)   &= \pi\ , \crm
b(-1/2) &= \frac{\sqrt{\pi}\,\Gamma(3/4)}{\Gamma(5/4)} = 2.3962804694711844\ldots\ , \crm
b(0)    &= 2\ , \crm
b(1/2)  &= \frac{\sqrt{\pi}\,\Gamma(1/4)}{\Gamma(3/4)} = 1.7480383695280799\ldots\ .
}
\ee
The normalized rotation kernel ${\cal R}(v)$ entering in the expression
of the amplitude $a_0$ (\ref{eq.appa0}), or (\ref{eq.appa0'}), which is
associated to a normalized intensity
\be
I(x,y) = \sum_n \gamma_n I_{\alpha_n}(x,y)
\ee
is then
\be
R(v) = \sum_n \frac{\gamma_n b(\alpha_n)}{V\sin i_\star}
\left(1-\left(\frac{v}{V\sin i_\star}\right)^2\right)^{(\alpha_n+1)/2}
\ .
\label{eq.Rv}
\ee

\section{RM signal measured by the iodine cell technique}
\label{sec.appdGconvR}
In this section, we compute an analytical expansion of $\vd$
(\ref{eq.vdconv}) modeling the RM signal measured by the 
iodine cell technique. The expansion is made possible if the subplanet
line profile $\Fpla$ and that of the nonrotating star ${\cal F}_0$ are
both Gaussian. In that case, if we denote ${\cal G}_{\beta_0}(v) = 
{\cal F}_0$, ${\cal G}_{\beta_p}(v) = \Fpla(v)$, and ${\cal R}(v)$ as the
rotation kernel, we have, on the one hand,
\be
\frac{d\Fstar(v)}{dv} = 
\left[\frac{d{\cal G}_{\beta_0}}{dv}*R\right](v)\ ,
\label{eq.Fstar'}
\ee
and, on the other hand,
\be
\left[\frac{d\Fstar}{dv}*\Fpla\right](v_p) = 
\left[\frac{d{\cal G}_{\beta_t}}{dv}*R\right](v_p)
\label{eq.numiodine}
\ee
with $\beta_t^2 = \beta_0^2+\beta_p^2$. Thus, both the numerator and the
denominator involve integrals of the form $\left(d\Gfit/dv\right)*R$.
Let us consider the case where ${\cal R} = {\cal R}_\alpha$ with
\be
{\cal R}_\alpha(v) = \frac{b(\alpha)}{V\sin i_\star}
\left(1-\left(\frac{v}{V\sin i_\star}\right)^2\right)^\eta
\ee
and $\eta = (\alpha+1)/2$. The expansion in series of
$(d\Gfit/dv)*{\cal R}_\alpha$ is obtained by expanding 
${\cal R}_\alpha$ in the vicinity of $v=0$. We have
\be
{\cal R}_\alpha(v) = \frac{b(\alpha)}{V\sin i_\star} \sum_{k=0}^{\infty}
\frac{(-\eta)_k}{k!} \left(\frac{v}{V\sin i_\star}\right)^{2k}\ ,
\ee
where $(-\eta)_k = 1$ if $k=0$, and $(-\eta)(-\eta+1)
\ldots(-\eta+k-1)$ otherwise, is the Pochhammer symbol. It should be
noted that if $\eta\in\mathbb{N}$, $(-\eta)_{\eta+1}=0$, then the
sum is finite and the expansion exact. In the following, we consider
only truncated sums up to an order $K$, i.e., $k\leq K$. We then have
\be
\EQM{
\bigg[\frac{d\Gfit}{dv}* & {\cal R}_\alpha\bigg](\vx) =
\frac{1}{\sqrt{2\pi}\sigma}\frac{b(\alpha)}{V\sin i_\star} 
\sum_{k=0}^K 
\Bigg(
\frac{(-\eta)_k}{k!} 
\crm & \times
\int_{-V\sin i_\star}^{V\sin i_\star} 
\left(\frac{v}{V\sin i_\star}\right)^{2k}
\frac{v-\vx}{\sigma^2} 
\exp\left(-\frac{(v-\vx)^2}{2\sigma^2}\right)
\,dv
\Bigg)\ .
}
\ee
We apply the change of variable $x=(v-\vx)/\sigma$, and set $x_1=-(V\sin
i_\star+\vx)/\sigma$ and $x_2=(V\sin i_\star-\vx)/\sigma$. We obtain
\be
\EQM{
\bigg[\frac{d\Gfit}{dv}*{\cal R}_\alpha\bigg](\vx) = & 
\frac{1}{\sqrt{2\pi}\sigma}\frac{b(\alpha)}{V\sin i_\star}
\sum_{k=0}^K
\Bigg(
\frac{(-\eta)_k}{k!} 
\crm & \times
\int_{x_1}^{x_2} 
\left(\frac{x\sigma+\vx}{V\sin i_\star}\right)^{2k}
x \e^{-x^2/2}
\,dx
\Bigg)\ .
}
\ee
The parenthesis inside the integral is then expanded which leads to
\be
\EQM{
\bigg[\frac{d\Gfit}{dv}* & {\cal R}_\alpha\bigg](\vx) =
\frac{1}{\sqrt{2\pi}\sigma}\frac{b(\alpha)}{V\sin i_\star}
\sum_{k=0}^K
\Bigg(
\frac{(-\eta)_k}{k!} 
\sum_{m=0}^{2k} \bpm 2k \\ m \epm
\crm & \times
\left(\frac{\vx}{V\sin i_\star}\right)^{2k-m}
\left(\frac{\sigma}{V\sin i_\star}\right)^m
\int_{x_1}^{x_2} 
x^{m+1} \e^{-x^2/2}
\,dx
\Bigg)\ .
}
\ee
The expression is easier to handle when the two sums are inverted. For
that, we introduce truncated hypergeometric functions of the form
\be
F_K(a,b,c;x) = \sum_{k=0}^K \frac{(a)_k(b)_k}{(c)_k} \frac{x^k}{k!}\ ,
\ee
and $Q_m(\eta; x)$ defined by
\be
Q_{2p}(\eta; x) =
\frac{(-\eta)_p}{p!} 
F_{K-p}\left(-\eta+p, p+\frac{1}{2}, \frac{1}{2}, x\right)\ ,
\ee
and
\be
\EQM{
Q_{2p+1}(\eta; x) = &
2x \frac{(-\eta)_{p+1}}{p!} 
\crm & \times
F_{K-p-1}\left(-\eta+p+1, p+\frac{3}{2}, \frac{3}{2}, x\right)\ .
}
\ee
With these notations, we have
\be
\EQM{
\bigg[\frac{d\Gfit}{dv}*{\cal R}_\alpha\bigg](\vx) = & 
\frac{1}{\sqrt{2\pi}\sigma}\frac{b(\alpha)}{V\sin i_\star}
\sum_{m=0}^{2K}
\Bigg(
Q_m\left(\eta; \frac{\vx}{V\sin i_\star}\right)
\crm & \times
\left(\frac{\sigma}{V\sin i_\star}\right)^m
\int_{x_1}^{x_2} 
x^{m+1} \e^{-x^2/2}
\,dx
\Bigg)\ ,
}
\label{eq.dGconvR}
\ee
where
\be
\int_0^{y} x^{m+1} \e^{-x^2/2}\,dx = 
({\rm sign}\,y)^m\, 2^{m/2} 
\gamma\left(1+\frac{m}{2}, \frac{y^2}{2}\right)\ ,
\ee
and $\gamma(a,x) = \int_0^x t^{a-1}\e^{-t}\,dt$ is the lower
incomplete gamma function.

The formula (\ref{eq.dGconvR}) gives the expansion of the numerator 
(\ref{eq.numiodine}) of $\vd$ (\ref{eq.vdconv}). Moreover, the
denominator, which is the integral of the square of $\Fstar'$ (\ref{eq.Fstar'}) 
can be computed by numerical integration using the same expansion. We
observe numerically that the convergence of the expansion of $\vd$, using 
Eq.~(\ref{eq.dGconvR}), is quite fast. In practice $K=4$ already gives
accurate results.

\bibliography{rossiter}

\end{document}